\documentclass[aps, prd, twocolumn, superscriptaddress, floatfix, longbibliography, nofootinbib]{revtex4-2}

\pdfoutput=1
\usepackage{amsmath}
\usepackage{amssymb}
\usepackage{graphicx}
\usepackage{booktabs}
\usepackage{multirow}
\usepackage{placeins}

\usepackage[colorlinks=true,breaklinks=true]{hyperref}
\usepackage[dvipsnames]{xcolor}
\definecolor{darkred}{rgb}{0.5,0,0}
\definecolor{darkblue}{rgb}{0,0,0.5}
\definecolor{firebrick}{rgb}{0.75,0.125,0.125}
\definecolor{darkgreen}{rgb}{0,0.5,0}
\hypersetup{urlcolor=darkblue,
	    citecolor=darkgreen,
	    linkcolor=firebrick}

\usepackage{orcidlink}
\usepackage{fontawesome5}
\newcommand{\repourl}{https://github.com/mbustama/nu-lri-earth}

\newcommand{\ie}{{\it i.e.}}

\newcommand{\Eg}{{\it E.g.}}

\newcommand{\eq}{Eq.}

\newcommand{\fig}{Fig.}

\providecommand{\Ref}{}
\AtBeginDocument{\renewcommand{\Ref}{Ref.}}
\newcommand{\Refs}{Refs.}

\newcommand{\equ}[1]{\eq~(\ref{equ:#1})}
\newcommand{\figu}[1]{\fig~\ref{fig:#1}}

\begin{document}

\title{Earth-sized tests of long-range neutrino interactions\,\href{\repourl}{\faGithub}}

\author{Dyer F.\ Cardenas\,\orcidlink{0009-0008-4071-8574}}
\email{dyer.cardenas@pucp.edu.pe}
\affiliation{Secci\'on F\'{\i}sica, Departamento de Ciencias, Pontificia Universidad Cat\'olica del Per\'u,\\Apartado 1761, Lima, Per\'u}

\author{Mauricio Bustamante\,\orcidlink{0000-0001-6923-0865}}
\email{mbustamante@nbi.ku.dk}
\affiliation{Niels Bohr International Academy, Niels Bohr Institute,\\University of Copenhagen, 2100 Copenhagen, Denmark}

\date{\today}

\begin{abstract}
Neutrinos may experience a new long-range force that distinguishes their flavors, mediated by an ultralight neutral boson of a new $U(1)'$ gauge symmetry and sourced by the electrons and nucleons of ordinary matter.
Atmospheric neutrinos that cross the Earth would then feel a flavor-dependent potential that alters their oscillations.
When the range of the force is about one Earth radius or less, this potential varies from point to point along each neutrino's path, since it comes mostly from the matter near the neutrino.
Earlier limits approximated it, either as constant along the path or as following the local density of matter.
We compute it exactly at every point of each path, for the Earth model of the IceCube analysis and over the distribution of heights at which the neutrinos are produced, and use it to analyze the 8-yr public IceCube DeepCore sample with the systematic uncertainties of that analysis.
For mediator masses from $10^{-15}$ to $10^{-9}$~eV, where the Earth alone sources the potential, we set upper limits at 90\% confidence level on the coupling of eight of 14 $U(1)'$ symmetries, those whose potential differs between $\nu_\mu$ and $\nu_\tau$.
A potential held at its value at the detector would weaken our limits by factors of about 1--3.
Below about $10^{-11}$--$10^{-10}$~eV, depending on the symmetry, our limits are stronger than those from tests of the equivalence principle.
\end{abstract}

\maketitle

\section{Introduction}
\label{sec:intro}

Do neutrinos feel a new long-range force, sourced by ordinary matter, that tells their flavors apart?
The idea of a long-range force coupled to a conserved charge goes back to \Ref~\cite{Lee:1955vk}, for baryon number; \Ref~\cite{Okun:1995dn} extended it to the number of each lepton flavor.
In the Standard Model, the differences between lepton-flavor numbers, such as $L_e - L_\mu$, can be gauged without anomalies~\cite{He:1990pn, He:1991qd, Foot:1990mn, Foot:1990uf, Foot:1994vd}.
Gauging one of them introduces a new neutral boson, $Z'$, of a $U(1)'$ symmetry~\cite{Langacker:2008yv, Heeck:2018nzc}.
For an ultralight $Z'$, with a mass $m'$ between $10^{-15}$ and $10^{-9}$~eV, the window that we study, the force has a range of about 200~m to 30 Earth radii.
Through this long-range interaction (LRI), the electrons, protons, and neutrons of the Earth create a potential that acts differently on each neutrino flavor, proportionally to the square of $g'$, the coupling of the $Z'$.

\begin{figure}[t]
    \centering
    \includegraphics[width=\columnwidth]{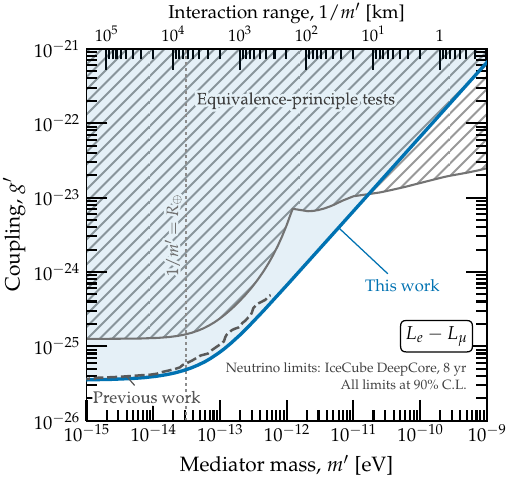}
    \caption{\textbf{\textit{Limits on the coupling of an $L_e - L_\mu$ interaction versus the mediator mass.}}
    Our limit uses the potential along each path through the Earth, with every oscillation and nuisance parameter left free in the fit (Sec.~\ref{sec:statistics}).
    The previous limit is from \Ref~\cite{Garg:2026gwx}, which holds the potential of the Earth constant along each path (the limit ends at the top edge of that reference's figure).
    The hatched region is excluded by tests of the equivalence principle~\cite{Wagner:2012ui, MICROSCOPE:2022doy} (Sec.~\ref{sec:bounds}).
    Figure~\ref{fig:all_limits} shows the limits on the other $U(1)'$ symmetries.}
    \label{fig:lri_limits}
\end{figure}

Figure~\ref{fig:lri_limits} shows present limits on $g'$ for $L_e - L_\mu$.
Tests of fifth forces and of the equivalence principle, with torsion balances~\cite{Schlamminger:2007ht, Adelberger:2009zz} and with a satellite in orbit~\cite{MICROSCOPE:2022doy}, look for a force that pulls differently on bodies of different composition.
They bound the coupling of the $Z'$ to electrons and nucleons~\cite{Blinnikov:1995kp}.
They constrain its coupling to neutrinos only indirectly, through the charges that each symmetry assigns to neutrinos and to matter.
Neutrino oscillations test the coupling to neutrinos directly, flavor by flavor: an LRI shifts the energy of each flavor by a different amount, changing how neutrinos oscillate.
For $L_e - L_\mu$, the neutrino limits are stronger than these bounds at ranges longer than about 12~km (\figu{lri_limits}).

Searches and forecasts for LRIs span solar and reactor neutrinos~\cite{Grifols:2003gy, Bandyopadhyay:2006uh, Gonzalez-Garcia:2006vic}, atmospheric neutrinos~\cite{Joshipura:2003jh, Samanta:2010zh, Khatun:2018lzs}, accelerator neutrinos~\cite{Heeck:2010pg, Davoudiasl:2011sz, Chatterjee:2015gta, Singh:2023nek, Agarwalla:2024ylc}, and high-energy astrophysical neutrinos~\cite{Bustamante:2018mzu, Agarwalla:2023sng}.
In our window of masses, the Earth is the only source of the potential that matters; the Moon, the Sun, and more distant matter add a negligible part.
The Earth is then both the source of the potential and the medium that the neutrinos cross.
Atmospheric neutrinos that arrive from below the horizon cross it along paths as long as its diameter, through the mantle and the core.
IceCube DeepCore~\cite{IceCube:2011ucd} has measured how these neutrinos oscillate, with high statistics~\cite{IceCubeCollaboration:2023wtb}, and made public its sample, together with the response of its detector.

Reference~\cite{Garg:2026gwx} used this sample to set the first limits on LRIs from DeepCore, for $L_e - L_\mu$ and $L_e - L_\tau$, and \Ref~\cite{Garg:2026rmy} extended them to a broad class of $U(1)'$ symmetries.
Like forecasts for long-baseline experiments~\cite{Singh:2023nek, Agarwalla:2024ylc} and studies of high-energy astrophysical neutrinos~\cite{Bustamante:2018mzu, Agarwalla:2023sng}, both evaluate the potential of the Earth at the detector and keep that value along the whole path.
Yet the potential changes along the path, and more so for mediators whose range is comparable to the radius of the Earth or shorter, since each stretch of the path then feels mostly the matter close to it (Sec.~\ref{sec:potential}).
In a global fit of oscillation data, \Ref~\cite{Coloma:2020gfv} instead rescales the potential of a contact interaction, which follows the local density, by one factor per source, electrons or neutrons, evaluated at the Earth's center.
That shape is accurate only when the range is short compared with the layers of the Earth.
References~\cite{Smirnov:2019cae, Wise:2018rnb} computed how the potential varies inside the Earth, but did not fit it to data.

In this work, we compute the potential at each point along the path of every neutrino, from the matter of a layered model of the Earth, and from each height in the atmosphere at which the neutrino can be produced.
Using this potential, we fit the DeepCore data, with the nuisance parameters, priors, and test statistic of the IceCube analysis.
We look in the data for the change that the LRI potential would make to the oscillations of $\nu_\mu$ that cross the Earth.
DeepCore is most sensitive to it near the minimum of their survival probability, at about 24~GeV (Sec.~\ref{sec:evolution}).
We study 14 $U(1)'$ symmetries~\cite{Agarwalla:2024ylc, Garg:2026rmy}.
For the eight whose potential differs between $\nu_\mu$ and $\nu_\tau$, we set upper limits on $g'$ across our window of masses; for the six whose potential acts on $\nu_e$ alone, the data set none, since such a potential barely changes $\nu_\mu$ oscillations (Sec.~\ref{sec:nue_only}).
To our knowledge, this is the first analysis of data that follows how the potential of the Earth changes along the path of each neutrino.

Figure~\ref{fig:lri_limits} summarizes our limits for one symmetry, $L_e - L_\mu$.
\textbf{Computing the potential along each path, instead of holding it at its value at the detector, makes our limit on $g'$ 1.2 times stronger at the longest ranges and 2--2.7 times stronger at ranges shorter than about 600~km.}
Our limit is 1.03--1.5 times stronger than the earlier DeepCore limit of \Ref~\cite{Garg:2026gwx} wherever both exist, but this comparison also reflects other differences between the two analyses (Sec.~\ref{sec:limits_lelmu}).
Among the other symmetries, those whose potential acts on $\nu_\mu - \nu_\tau$ have the strongest limits, by a factor of about 1.45 once their charges are accounted for (\figu{all_limits}), since their potential separates $\nu_\mu$ from $\nu_\tau$ twice as much.

Section~\ref{sec:lri_earth} describes the potential of the Earth and how it changes the oscillations of atmospheric neutrinos.
Section~\ref{sec:deepcore} describes the DeepCore sample and our analysis of it.
Section~\ref{sec:results} presents our limits, which Sec.~\ref{sec:summary} summarizes.
The appendices contain the closed form of the potential (Appendix~\ref{app:lri_potential}), the atmospheric neutrino flux (Appendix~\ref{app:flux}), the parameters of our fit (Appendix~\ref{app:parameters}), our fit without an LRI (Appendix~\ref{app:fit_no_lri}), and a test of the coverage of our limits (Appendix~\ref{app:coverage}).

\section{Long-range interactions inside the Earth}
\label{sec:lri_earth}

\subsection{$U(1)'$ symmetries and their textures}
\label{sec:symmetries}

\smallskip
\textit{Symmetries.}---The Standard Model has accidental global symmetries associated with baryon number, $B$, and with the three lepton numbers, $L_e$, $L_\mu$, and $L_\tau$.
Some combinations of them can be gauged without anomalies, with the particle content of the Standard Model or after adding right-handed neutrinos~\cite{He:1990pn, He:1991qd, Foot:1990mn, Foot:1990uf, Foot:1994vd, Araki:2012ip, Allanach:2018vjg}.
Each such $U(1)'$ symmetry brings a new neutral vector boson, $Z'$, of mass $m'$.
Its interaction with the fermions of ordinary matter and with neutrinos is~\cite{Agarwalla:2024ylc}
\begin{multline}
 \label{equ:lagrangian}
 \mathcal{L}_{Z'} = -g' \Big( \sum_{f = u, d, e} a_f\, \bar f \gamma^\mu f \\
 + \sum_{\alpha = e, \mu, \tau} b_\alpha\, \bar\nu_\alpha \gamma^\mu P_L \nu_\alpha \Big) Z'_\mu \;,
\end{multline}
where $g'$ is the coupling, $a_u$, $a_d$, and $a_e$ are the $U(1)'$ charges of the up quark, the down quark, and the electron, $b_\alpha$ is the charge of $\nu_\alpha$, and $P_L$ is the left-handed chirality projector.
A proton has charge $a_p = 2a_u + a_d$ and a neutron has charge $a_n = 2a_d + a_u$.

Table~\ref{tab:symmetries} lists the 14 flavor-dependent symmetries that we study, with the texture and the charge factor of each, defined below.
They are those of \Refs~\cite{Agarwalla:2024ylc, Garg:2026rmy}, which follow the list of \Refs~\cite{Coloma:2020gfv, Davoudiasl:2011sz, Araki:2012ip, delaVega:2021wpx}.
References~\cite{Ma:1997nq, Lee:2010hf, Kownacki:2016pmx, Farzan:2015doa} discuss models that realize some of them.
(Flavor-universal symmetries, such as $B - L$, do not affect oscillations.)
In the symmetry $B_y + L_\mu + L_\tau$, the charge of the quarks is $B_y = B_1 - y B_2 - (3 - y) B_3$, where $B_i$ is the baryon number of generation $i$ and $y$ is a free constant~\cite{Farzan:2015doa, Coloma:2020gfv}.
Our results do not depend on $y$, since ordinary matter holds only quarks of the first generation, whose charge, 1/3, is the same as under $B$.
Ordinary matter carries no $L_\mu - L_\tau$ charge.
Under this symmetry, neutrinos feel matter through the mixing of the $Z'$ with the $Z$ boson, which couples them to neutrons~\cite{Holdom:1985ag, Babu:1997st, Heeck:2010pg, Joshipura:2019qxz}.
For it, as in \Refs~\cite{Agarwalla:2024ylc, Garg:2026rmy}, each neutron carries the charge $a_n = e/(\sin\theta_W \cos\theta_W)$, where $e$ is the electric charge and $\theta_W$ the weak mixing angle; for $\alpha = 1/137$ and $\sin^2\theta_W = 0.231$, $a_n = 0.72$.
The coupling $g'$ of \equ{potential} is replaced by the effective coupling $G' = [g'(\xi - \sin\theta_W\, \chi)]^{1/2}$, where $\chi$ is the kinetic mixing between $Z'$ and $Z$ and $\xi$ the angle between their mass and gauge eigenstates.
For $L_\mu - L_\tau$, our limits are on $G'$.

\begin{table}[t]
 \caption{\label{tab:symmetries}\textbf{\textit{The $U(1)'$ symmetries, the texture of their potential, and their charge factor.}}
 The charges are those of \Refs~\cite{Agarwalla:2024ylc, Garg:2026rmy}; $k$ is that of \equ{charge_factor}, evaluated with $\Psi_n/\Psi_e = 1.017$, the ratio of the neutron and electron densities at the detector; our fits use $k(\mathbf{r})$ along each path.
 For $L_\mu - L_\tau$, $k$ uses the charge of the neutrons under $Z$--$Z'$ mixing (Sec.~\ref{sec:symmetries}), and both signs are allowed.
 At equal $g'^2 k$, symmetries with the same texture have nearly the same potential, and so nearly the same limit on $g'\sqrt{|k|}$.}
 \begin{ruledtabular}
 \begin{tabular}{l l r}
 Symmetry & Potential acts on & Charge factor, $k$ \\
 \hline
 $B - 3L_\mu$ & $\nu_\mu$ & $-6.05$ \\
 $L - 3L_\mu$ & $\nu_\mu$ & $-3.00$ \\
 $B - 3L_\tau$ & $\nu_\tau$ & $-6.05$ \\
 $L - 3L_\tau$ & $\nu_\tau$ & $-3.00$ \\
 $L_\mu - L_\tau$ & $\nu_\mu - \nu_\tau$ & $\pm 0.73$ \\
 $B - L_e - 2L_\tau$ & $\nu_\mu - \nu_\tau$ & $+1.02$ \\
 $L_e - L_\mu$ & $\nu_e - \nu_\mu$ & $+1.00$ \\
 $L_e - L_\tau$ & $\nu_e - \nu_\tau$ & $+1.00$ \\
 $B - 3L_e$ & $\nu_e$\footnotemark[1] & $+2.95$ \\
 $L - 3L_e$ & $\nu_e$\footnotemark[1] & $+6.00$ \\
 $B - \frac{3}{2}(L_\mu + L_\tau)$ & $\nu_e$\footnotemark[1] & $+3.03$ \\
 $L_e - \frac{1}{2}(L_\mu + L_\tau)$ & $\nu_e$\footnotemark[1] & $+1.50$ \\
 $L_e + 2L_\mu + 2L_\tau$ & $\nu_e$\footnotemark[1] & $-1.00$ \\
 $B_y + L_\mu + L_\tau$ & $\nu_e$\footnotemark[1] & $-2.02$ \\
 \end{tabular}
 \end{ruledtabular}
 \footnotetext[1]{Weak or no sensitivity; see Sec.~\ref{sec:nue_only}.}
\end{table}

\smallskip
\textit{Potential.}---A neutrino at a distance $d$ from a fermion $f$ feels, through the exchange of a $Z'$, the Yukawa potential $g'^2 b_\alpha a_f\, e^{-m' d}/(4\pi d)$~\cite{Agarwalla:2024ylc}.
The potential on $\nu_\alpha$ at $\mathbf{r}$ is the sum over all the electrons, protons, and neutrons,
\begin{align}
 V_\alpha(\mathbf{r}) &= g'^2\, b_\alpha \sum_{f = e, p, n} a_f\, \Psi_f(\mathbf{r}) \;, \label{equ:potential} \\
 \Psi_f(\mathbf{r}) &= \int d^3 r'\, n_f(\mathbf{r}')\, \frac{e^{-m' |\mathbf{r} - \mathbf{r}'|}}{4\pi |\mathbf{r} - \mathbf{r}'|} \;, \label{equ:psi}
\end{align}
where $n_f$ is the number density of fermion $f$.
An antineutrino, whose charge is $-b_\alpha$, feels $-V_\alpha$.
Matter farther away than a few interaction ranges, $1/m'$, contributes little, so the range determines which part of the Earth creates the potential at each point of a path (Sec.~\ref{sec:potential}).

\smallskip
\textit{Textures.}---Only the differences between $V_e$, $V_\mu$, and $V_\tau$ affect oscillations, since a potential common to the three flavors adds a multiple of the identity to the Hamiltonian of \equ{hamiltonian} below, which changes only a global phase that does not affect oscillation probabilities.
We write the neutrino charges as $b_\alpha = c\, q_\alpha + b_0$, where $c$ and $b_0$ are numbers and $q = (q_e, q_\mu, q_\tau)$ is one of six textures: $(1, 0, 0)$, $(0, 1, 0)$, $(0, 0, 1)$, $(1, -1, 0)$, $(1, 0, -1)$, and $(0, 1, -1)$.
Since $b_0$ is common to the three flavors, the potential that affects oscillations acts only on the flavors with $q_\alpha \neq 0$. Each texture has a flavor with $q_\alpha = 0$, whose neutrino charge is $b_0$, and one with $q_\alpha = 1$, whose neutrino charge is $c + b_0$.
\Eg, for $L - 3L_e$, with texture $(1, 0, 0)$ and charges $(b_e, b_\mu, b_\tau) = (-2, 1, 1)$, this gives $c = -3$ and $b_0 = 1$.
Electrons and protons have equal densities in neutral matter, so $\Psi_p = \Psi_e$.
The part of the potential that affects oscillations is then
\begin{align}
 \mathbf{V}_{\rm LRI}(\mathbf{r}) &= g'^2\, k(\mathbf{r})\, \Psi_e(\mathbf{r})\, {\rm diag}(q_e, q_\mu, q_\tau) \;, \label{equ:texture} \\
 k(\mathbf{r}) &\equiv c \left[ a_e + a_p + a_n \frac{\Psi_n(\mathbf{r})}{\Psi_e(\mathbf{r})} \right] \;, \label{equ:charge_factor}
\end{align}
where $k$ is the charge factor of the symmetry (Table~\ref{tab:symmetries}).

For the symmetries without baryon number, except $L_\mu - L_\tau$, electrons are the only sources, so $k = c\, a_e$ is constant.
Their potential depends on $g'$ and on their charges only through $g'^2 k$, so two of them with the same texture and the same sign of $k$ have the same limit on $g'^2 |k|$.
For the other symmetries, $k$ changes along the path, since the ratio of neutrons to electrons in the matter that creates the potential, $\Psi_n/\Psi_e$, lies between that of the mantle, 1.02, and that of the core, 1.15.
References~\cite{Agarwalla:2024ylc, Garg:2026rmy} group the symmetries by texture in the same way, but for a potential held constant along the path.
Only the $\nu_e$ texture contains symmetries with both signs of $k$.
For it, the LRI potential either adds to the standard matter potential or subtracts from it (Sec.~\ref{sec:evolution}).

\subsection{The potential of the Earth}
\label{sec:potential}

We compute the matter potential using the same Earth model that the DeepCore analysis by IceCube uses~\cite{IceCube:2018ikn}: a simplified version of the Preliminary Reference Earth Model (PREM)~\cite{Dziewonski:1981xy}, with 12 layers of constant density.
(The full PREM, whose density varies within each layer, changes the expected number of events in each bin by $\lesssim 0.1\%$.)
Its electron fraction, the number of electrons per nucleon, is 0.4656 in the core, inner and outer, and 0.4957 in every layer above it.
The same model gives the standard matter potential (Sec.~\ref{sec:evolution}).
In the spherical Earth we assume, $\Psi_f$ depends only on the distance $r$ from the center.
Its angular integral can then be done analytically, which leaves a radial integral with a closed form in each layer, valid at every interaction range that we study (Appendix~\ref{app:lri_potential}).
Above the ground, $\Psi_f$ falls with height but does not vanish, so the potential also acts on the stretch of the path of the neutrino in the atmosphere. We leave out the matter of the air, both as a source of the LRI potential and in the standard matter potential.
For $L_e - L_\mu$, with $g'$ at our 90\%~C.L.\ upper limit (Sec.~\ref{sec:limits_lelmu}), including it changes the expected number of events in each bin by less than $10^{-7}$, for $m' = 10^{-15}$--$10^{-9}$~eV.

\begin{figure*}[t]
    \centering
    \includegraphics[width=\textwidth]{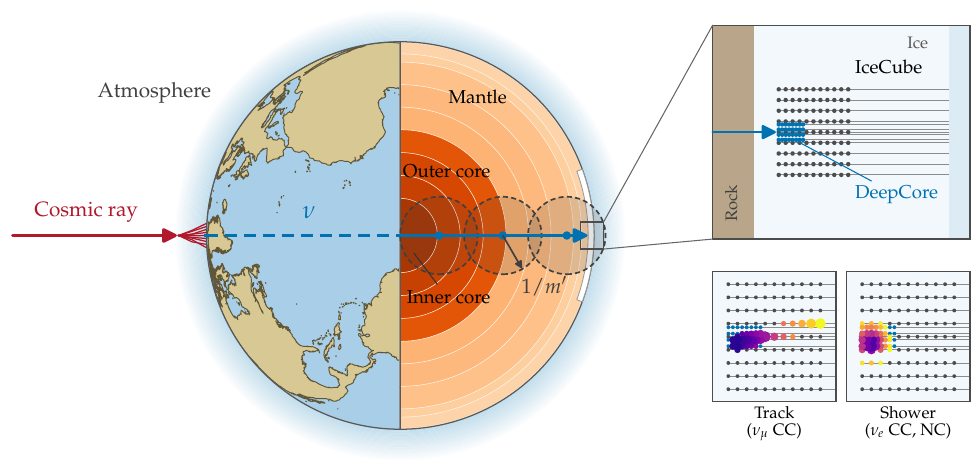}
    \caption{\textbf{\textit{An atmospheric neutrino crossing the Earth, with a long-range interaction.}}
    \textit{Left half:} the surface of the Earth, behind which the path is hidden (dashed).
    \textit{Right half:} the interior, in the 12 layers of the Earth model, shaded by density.
    The potential at each of three points of the path comes mostly from the matter within a few interaction ranges, $1/m'$, of it.
    Near the surface, part of that region holds no matter.
    \textit{Lower right:} stylized events, with hits sized by charge and colored by arrival time; the most shower-like events are not in the sample.
    As the neutrino moves, so does the matter that creates its potential.
    Not to scale; see Sec.~\ref{sec:potential} for details.}
    \label{fig:lri_concept}
\end{figure*}

Figure~\ref{fig:lri_concept} shows the setup.
A neutrino produced in the atmosphere crosses the Earth on its way to DeepCore at the South Pole.
The circles in the figure have radius $1/m'$, the interaction range, around three points of its path.
Inside uniform matter, the matter within one range of a point creates 26\% of the potential there; the matter within three ranges creates 80\%.
So, for ranges shorter than the Earth's radius, $R_\oplus = 6371$~km, each part of the path feels a different part of the Earth.

\begin{figure*}[t]
    \centering
    \includegraphics[width=\textwidth]{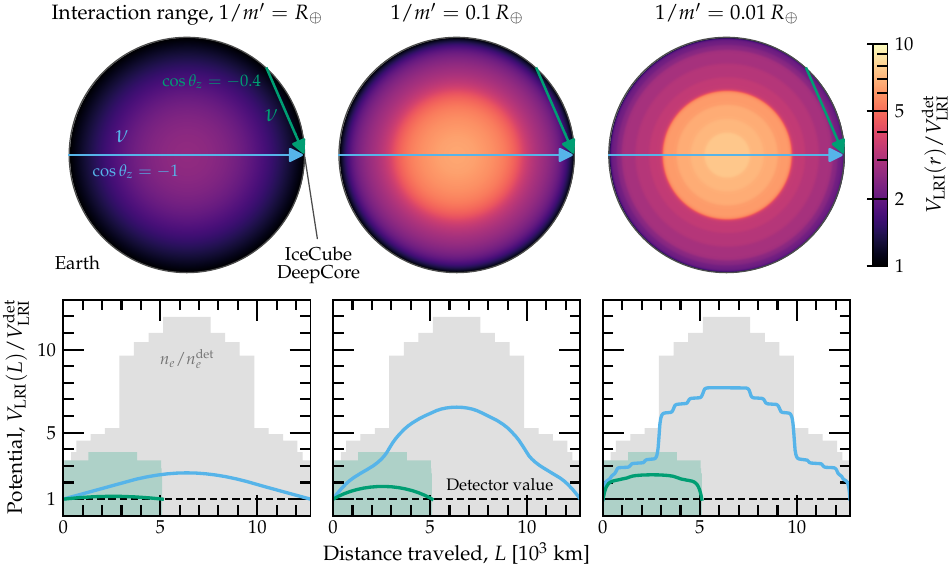}
    \caption{\textbf{\textit{The long-range potential inside the Earth, relative to its value at the detector.}}
    \textit{Top:} in a cross-section of the Earth, for $L_e - L_\mu$, whose source is the electrons, with two paths to IceCube.
    \textit{Bottom:} along the two paths, with the electron density along each, relative to that at the detector, shaded in gray for the path through the center and in green for the other.
    The potential approaches that density ratio only when the range is well below 1~km, the distance from the detector to the denser rock below it.
    See Sec.~\ref{sec:potential} for details.}
    \label{fig:vlri_trajectory}
\end{figure*}

Figure~\ref{fig:vlri_trajectory} shows the potential inside the Earth relative to its value at the detector, for three interaction ranges.
For ranges much longer than $R_\oplus$, the whole Earth contributes to the potential at every point.
Each piece of matter then contributes in inverse proportion to its distance, so the potential is larger from matter that is closer.
For $m' \to 0$, it is 1.76 times larger at the center, which lies within one Earth radius of all the matter, than at the detector.
For ranges much shorter than the scale over which the density changes, only the matter near a point contributes.
There, $\Psi_f$ tends to $n_f/m'^2$, the contact limit, so the potential follows the local density and jumps at the boundaries between PREM layers.
Between these two limits, the potential follows the radial structure of the Earth, smoothed over the interaction range.

The detector lies at a depth of 2~km, in the top layer of the model, which is water, 1~km above the denser crust.
For $1/m' = 0.2$~km, the shortest range that we study, the potential at the center of the Earth is 11.9 times that at the detector, close to the ratio of the electron densities of the inner core and the water, 12.
At longer ranges, the denser rock below the detector adds to the potential there, so this ratio falls, to 7.7 for $1/m' = 64$~km.

\subsection{Comparison with earlier approaches}
\label{sec:approaches}

The constant potential of earlier analyses (Sec.~\ref{sec:intro}) misses how the potential changes along each path (\figu{vlri_trajectory}).
Section~\ref{sec:limits_lelmu} compares our limits with those of the constant potential, with everything else unchanged.

Earlier analyses also reach mediator masses far below our window, where matter beyond the Earth sources part of the potential~\cite{Bustamante:2018mzu, Coloma:2020gfv, Agarwalla:2024ylc, Garg:2026gwx, Garg:2026rmy}.
Down to a few times $10^{-18}$~eV, the Earth still sources most of it; by our estimate, the Sun adds 1\% at $10^{-17}$~eV.
With the Earth alone, our limit on $g'$ for $m' \to 0$ would be only about 1.5\% below its value at $10^{-15}$~eV, yet the potential still varies along each path (Sec.~\ref{sec:potential}).
For lighter mediators, the Sun, the Milky Way, and the cosmological distribution of matter take over~\cite{Bustamante:2018mzu}.
Their potential is the same at every point of every path inside the Earth.
Once the range far exceeds the distance to the Sun, the change of the Earth's potential along a path alters the total by 3\% at most, by our estimate.
A constant potential is then accurate, so we leave these masses to the earlier analyses.

\subsection{Flavor evolution}
\label{sec:evolution}

Along a path, the flavor state of a neutrino of energy $E$ evolves under the Hamiltonian, in the flavor basis,
\begin{multline}
 \label{equ:hamiltonian}
 \mathbb{H}(E, \mathbf{r}) = \frac{1}{2E}\, U\, {\rm diag}(0, \Delta m^2_{21}, \Delta m^2_{31})\, U^\dagger \\
 + \sqrt{2} G_F\, n_e(\mathbf{r})\, {\rm diag}(1, 0, 0) + \mathbf{V}_{\rm LRI}(\mathbf{r}) \;.
\end{multline}
Here, $U$ is the Pontecorvo--Maki--Nakagawa--Sakata (PMNS) mixing matrix~\cite{Pontecorvo:1957qd, Pontecorvo:1967fh, Maki:1962mu}, $\Delta m^2_{ij} = m_i^2 - m_j^2$ are the differences of the squared neutrino masses, and $\sqrt{2} G_F n_e$ is the standard matter potential~\cite{Wolfenstein:1977ue, Mikheyev:1985zog}, where $G_F$ is the Fermi constant.
The LRI potential, $\mathbf{V}_{\rm LRI}$, is that of \equ{texture}.
For antineutrinos, $U$ is replaced by $U^*$ and both potentials change sign.
We compute the evolution with Magnus~\cite{Bustamante:2026yce}, from the point of production to the detector, with both evaluated at every point of the path.

Atmospheric neutrinos come from the decays of pions, kaons, and muons in the air showers that cosmic rays create in the atmosphere~\cite{Gaisser:2002jj}.
We compute their flux with MCEq~\cite{Fedynitch:2015zma}, which solves the equations of these showers along each path through the atmosphere, with the SIBYLL-2.3d model of hadronic interactions~\cite{Riehn:2019jet} and the Global Spline Fit (GSF) model of cosmic rays~\cite{Dembinski:2017zsh, Fedynitch:2026ugq}.
Besides the flux at the ground, MCEq gives the flux produced at each height.
The DeepCore analysis uses the Honda flux~\cite{Honda:2015fha}; Appendix~\ref{app:flux} compares the two.
The length of a path, and the part of it in the air, where the LRI potential acts, depend on the height at which the neutrino is produced.
In our flux, the median production height at 10--30~GeV is 13--25~km for $\nu_\mu$ and 10--22~km for $\nu_e$, depending on direction.
We fold the oscillation probability with the distribution of production heights,
\begin{multline}
 \label{equ:prob_folded}
 \bar P_{\alpha\beta}(E, \theta_z) = \frac{1}{\Phi_\alpha(E, \theta_z)} \\
 \times \int dh\, \frac{d\Phi_\alpha}{dh}(E, \theta_z, h)\, P_{\alpha\beta}(E, \theta_z, h) \;,
\end{multline}
where $\theta_z$ is the zenith angle of the direction from which the neutrino arrives at the detector, with $\cos\theta_z = -1$ for neutrinos that arrive from directly below, $d\Phi_\alpha/dh$ is the flux of $\nu_\alpha$ produced at height $h$, and $\Phi_\alpha$ is the total flux.
We do this separately for $\nu_e$, $\bar\nu_e$, $\nu_\mu$, $\bar\nu_\mu$, $\nu_\tau$, and $\bar\nu_\tau$, each with its own flux and distribution of production heights.
The flux of $\nu_\tau$ and $\bar\nu_\tau$ is much smaller, below $10^{-3}$ of that of $\nu_\mu$ and $\bar\nu_\mu$.
Since neutrinos produced at different heights do not interfere, the expected events of \equ{events} later depend on the probability only through this flux-weighted average.
Using a single production height of 20~km---the choice of IceCube's analysis code~\cite{IceCube:2018ikn}---would change the expected events by up to 0.5\%.

At the highest energies of the sample, some of the neutrinos that cross the Earth are absorbed.
We include this through the transmission $\mathcal{T}(E, \theta_z)$ of \equ{events}, the fraction of the flux entering the Earth that reaches the detector.
If scattering only removed neutrinos, it would be $e^{-\tau}$, with the optical depth $\tau = N_A\, \sigma(E)\, X(\theta_z)$, where $N_A$ is Avogadro's number, $\sigma$ is the charged- plus neutral-current cross section per nucleon, and $X$ is the column density, \ie, the mass per unit area, along the path.
However, a neutrino that scatters through the neutral current continues at a lower energy, so the loss is about 10\% smaller.
We compute $\mathcal{T}$ by solving the transport equation of the flux, which includes both effects, along each path through the Earth model, for an incoming atmospheric-neutrino flux that falls as $E^{-3.7}$ and the CSMS cross sections of deep-inelastic neutrino-nucleon scattering~\cite{Cooper-Sarkar:2011jtt}.
Since we take these cross sections to be the same for all flavors, $\mathcal{T}$ multiplies the oscillation probability.
Absorption removes 0.5\% of the neutrinos that cross the center of the Earth at 100~GeV, and less than 1\% of our events.
The simulated events of the data release do not include it, nor does the DeepCore analysis~\cite{IceCubeCollaboration:2023wtb} mention it.

\begin{figure}[t]
    \centering
    \includegraphics[width=\columnwidth]{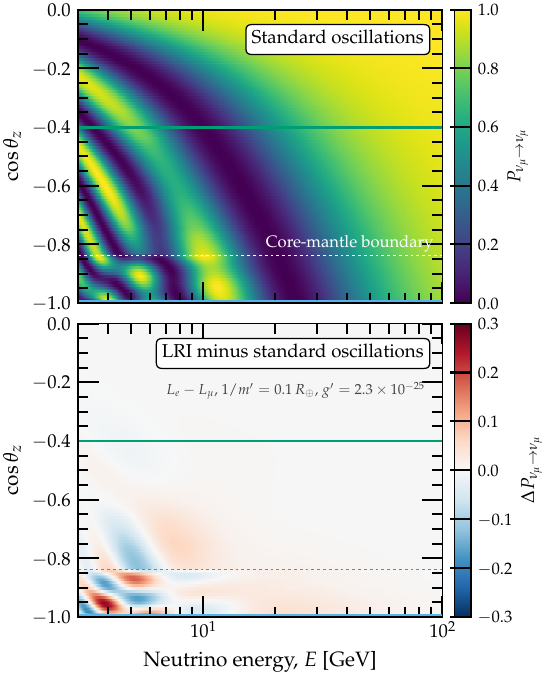}
    \caption{\textbf{\textit{Muon-neutrino survival probability versus energy and direction, without and with a long-range interaction.}}
    For neutrinos produced 20~km above the surface, the best-fit oscillation parameters of the DeepCore analysis, and the benchmark interaction, close to the expected limit at its interaction range.
    The colored lines mark the paths of Figs.~\ref{fig:vlri_trajectory} and~\ref{fig:oscillation_probabilities}.
    The largest changes lie below 10~GeV, in directions that cross the core.
    See Sec.~\ref{sec:evolution} for details.}
    \label{fig:oscillograms}
\end{figure}

Figure~\ref{fig:oscillograms} shows the survival probability of $\nu_\mu$ versus energy and direction, without and with an LRI, for the best-fit oscillation parameters of the DeepCore analysis~\cite{IceCubeCollaboration:2023wtb}.
As a benchmark, we take $L_e - L_\mu$ with $1/m' = 0.1\,R_\oplus$ and $g' = 2.34 \times 10^{-25}$, close to our expected 90\%~C.L.\ limit there.
The changes are largest where the potential is several times its value at the detector, along the paths that cross the core (\figu{vlri_trajectory}), and where the probability oscillates fastest, below 10~GeV.
These low-energy changes barely affect the expected events (Sec.~\ref{sec:events}).

\begin{figure}[t]
    \centering
    \includegraphics[width=\columnwidth]{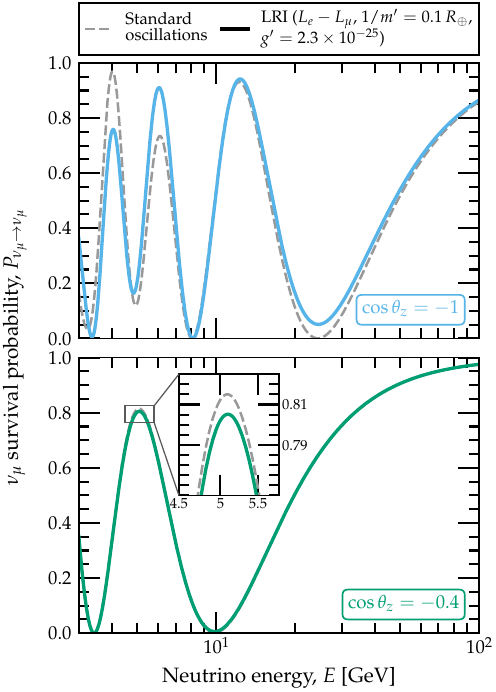}
    \caption{\textbf{\textit{Muon-neutrino survival probability versus energy, along two paths.}}
    For the configuration of Fig.~\ref{fig:oscillograms}, along the paths of Fig.~\ref{fig:vlri_trajectory}.
    Along the path that crosses the center, the interaction shifts the maxima below 10~GeV and lifts the minimum near 24~GeV above zero.
    Along the path through the mantle, where the potential is weaker, it changes the probability only slightly.
    See Sec.~\ref{sec:evolution} for details.}
    \label{fig:oscillation_probabilities}
\end{figure}

Figure~\ref{fig:oscillation_probabilities} shows the survival probability along two paths.
Along the path through the center of the Earth, $\cos\theta_z = -1$, the LRI keeps the probability from reaching zero at its minimum near 24~GeV.
The LRI potential separates $\nu_\mu$ from $\nu_\tau$, which reduces their effective mixing, the more so the larger the potential is compared to $\Delta m^2_{31}/(2E)$.
The potential does not depend on energy, whereas $\Delta m^2_{31}/(2E)$ falls with it, so the effect grows with energy.
Averaged along this path, the potential of the benchmark equals $\Delta m^2_{31}/(2E)$ at $E = 130$~GeV, and is about a fifth of it at the minimum near 24~GeV.
It is most visible at the minimum, where standard oscillations alone would make the probability vanish.
Along $\cos\theta_z = -0.4$, which crosses only the mantle, the potential is weaker and the changes are small (inset).
The filled minimum, in steeply up-going $\nu_\mu$ at 20--25~GeV, is what makes DeepCore sensitive to the LRI (Sec.~\ref{sec:events}).

\subsection{Loss of sensitivity to a potential on $\nu_e$ alone}
\label{sec:nue_only}

Six of our symmetries have equal charges for $\nu_\mu$ and $\nu_\tau$, so their potential acts on $\nu_e$ alone, with the texture $(1, 0, 0)$ (Table~\ref{tab:symmetries}).
They are $B - 3L_e$, $L - 3L_e$, $L_e - \frac{1}{2}(L_\mu + L_\tau)$, and $B - \frac{3}{2}(L_\mu + L_\tau)$, with $k > 0$, and $L_e + 2L_\mu + 2L_\tau$ and $B_y + L_\mu + L_\tau$, with $k < 0$.
In every other texture, the potential differs between $\nu_\mu$ and $\nu_\tau$ and partly fills the minimum of the $\nu_\mu$ survival probability near 24~GeV, on which the sensitivity of DeepCore rests (Sec.~\ref{sec:evolution}).
A potential equal for $\nu_\mu$ and $\nu_\tau$ leaves their mixing, and so the minimum, unchanged.

A potential on $\nu_e$ alone changes the survival of $\nu_\mu$ only through the mixing of $\nu_e$ with $\nu_\mu$, which, at the energies of DeepCore, is set by $\theta_{13}$.
Once the potential exceeds $\Delta m^2_{31}/(2E)$, it suppresses this mixing, as the standard matter potential does at high energies.
Then $\nu_e$ decouples: it no longer mixes with $\nu_\mu$ and $\nu_\tau$.
Beyond that point, larger couplings $g'$ barely change the oscillations.
So the test statistic of our fit, $\Delta\chi^2$ (Sec.~\ref{sec:statistics}), stops growing with the coupling and levels off once $\nu_e$ has decoupled.
This level lies below the threshold of our upper limits, so these symmetries receive none (Sec.~\ref{sec:limits_all}).

For $L_e + 2L_\mu + 2L_\tau$ and $B_y + L_\mu + L_\tau$, whose potential on $\nu_e$ is opposite in sign to the standard matter potential, $\Delta\chi^2$ first rises above this level over a range of couplings, so the data can exclude a bounded interval of couplings (Appendix~\ref{app:more_limits}).
Reference~\cite{Garg:2026rmy} left these six symmetries out of its analysis, since a potential on $\nu_e$ alone has little effect on the $\nu_\mu$ of its sample; we include them.

\section{Analysis of DeepCore data}
\label{sec:deepcore}

\subsection{Data sample}
\label{sec:sample}

We use the public data release~\cite{IceCubeDeepCoreRelease:2025} of the oscillation analysis of the IceCube Collaboration with DeepCore~\cite{IceCubeCollaboration:2023wtb}, the more densely instrumented inner part of IceCube~\cite{IceCube:2011ucd}.
Its data were taken over 8~years, between 2011 and 2019, with a livetime of 2724~days (7.5~years).
The events are binned in reconstructed energy, $E_{\rm reco}$, and reconstructed direction, $\cos\theta_{\rm reco}$: 10 logarithmic energy bins from 6.31 to 158.49~GeV, the last one twice as wide, and 10 direction bins from $-1$ to 0.1.
The release also provides the simulated events, with their true and reconstructed energies and directions, from which we compute the expected events (Sec.~\ref{sec:events}).

A particle-identification (PID) score separates the events further into two channels.
The charged-current interactions of $\nu_\mu$ make a muon, which leaves a long track; the other interactions make mostly showers (\figu{lri_concept}).
The Mixed channel, with scores of 0.55--0.75, holds mostly $\nu_\mu$ charged-current events, with about 30\% of events of other flavors and interactions.
The Tracks channel, with scores of 0.75--1, has a $\nu_\mu$ charged-current purity of 94\%.
In all, the sample has 200 bins and 21,914 events.
The Tracks channel measures the $\nu_\mu$ disappearance of Sec.~\ref{sec:evolution} most cleanly; the Mixed channel, with a similar number of events, adds statistical power.

\begin{figure}[t]
    \centering
    \includegraphics[width=\columnwidth]{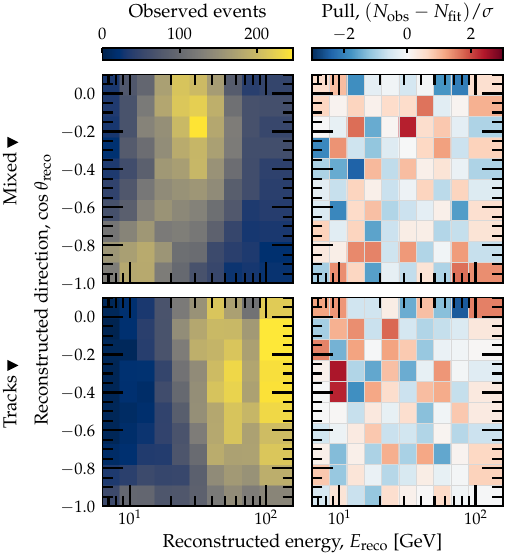}
    \caption{\textbf{\textit{Observed events of the DeepCore sample, and pulls against our fit without an LRI.}}
    \textit{Left:} observed events in each bin of reconstructed energy and direction, for the two PID channels, Mixed and Tracks.
    \textit{Right:} their pulls against our no-LRI fit, defined in Sec.~\ref{sec:statistics}.
    For this figure only, the fit fixes $\theta_{12}$, $\theta_{13}$, $\Delta m^2_{21}$, and $\delta_{\rm CP}$ at the values of the DeepCore analysis. In our full fits, these parameters are free.
    See Secs.~\ref{sec:sample} and~\ref{sec:statistics}, and Appendix~\ref{app:fit_no_lri}, for details.}
    \label{fig:data_fit}
\end{figure}

Figure~\ref{fig:data_fit} shows the observed events (left) and their pulls (right), which we discuss in Sec.~\ref{sec:statistics}.
At $E_{\rm reco} = 15$--37~GeV, the Tracks channel holds fewer steeply up-going events ($\cos\theta_{\rm reco} < -0.78$) than near-horizontal ones ($\cos\theta_{\rm reco} > -0.12$).
Without oscillations, the up-going events would outnumber the near-horizontal ones by factors of 1.3--2.6.
Standard oscillations remove two thirds of the up-going events.
The deficit comes from $\nu_\mu$ disappearance, deepest near 24~GeV along the steepest paths (Sec.~\ref{sec:evolution}); an LRI would partly refill it.

\subsection{Expected events}
\label{sec:events}

Each simulated event $i$ of the release carries a weight $w_i$, which, multiplied by the flux and by the livetime, $T_{\rm live}$, gives the contribution of that event to the expected number of events.
Like the observed events, the simulated events are placed in bins by their reconstructed energy and direction, and in the Mixed or Tracks channel by their PID score.
The expected number of events in a bin sums the contributions of its simulated events,
\begin{multline}
 \label{equ:events}
 N = A_{\rm eff} \sum_i f_{g_i}\, T_{\rm live}\, w_i\, x_i\, \mathcal{T}(E_i, \theta_i) \\
 \times \sum_\alpha \Phi_\alpha(E_i, \theta_i)\, \bar P_{\alpha\beta_i}(E_i, \theta_i) + s_\mu N_\mu \;,
\end{multline}
where $E_i$ and $\theta_i$ are the true energy and zenith angle of event $i$, which set its flux and probability, and $\beta_i$ is the flavor of the neutrino that interacts.
Here, $\Phi_\alpha$ is the flux of $\nu_\alpha$ at production, with the flux nuisance parameters applied, $\bar P_{\alpha\beta}$ is the probability of \equ{prob_folded}, $\mathcal{T}$ is the transmission through the Earth (Sec.~\ref{sec:evolution}), and $x_i$ reweights the cross sections.

The detector nuisance parameters enter through the hypersurfaces of the data release, factors that depend on these parameters and scale the expected events in each bin.
The release gives them separately for three groups of interactions: the neutral-current interactions of all flavors together with the charged-current interactions of $\nu_e$, those of $\nu_\mu$, and those of $\nu_\tau$.
Each simulated event is scaled by the hypersurface of its group, $f_{g_i}$.
The groups are set by the true interaction, not by the PID score, so every bin holds simulated events of all three.
The factor $A_{\rm eff}$ normalizes all the neutrino events, and $N_\mu$ is the template of atmospheric muons of the release, scaled by $s_\mu$.
The LRI enters only through $\bar P$, before detection; everything after it is as in the DeepCore analysis.

We use the 18 nuisance parameters of the DeepCore analysis, with its formulas and priors~\cite{IceCubeCollaboration:2023wtb} (Appendix~\ref{app:parameters}).
Five describe the detector: the efficiency of the optical modules, two parameters of the refrozen ice in the holes around them, and the absorption and scattering of the bulk ice.
Seven describe the flux: a change of its spectral index and six parameters of hadron production in air showers~\cite{Barr:2006it}.
Four describe the cross sections: the axial masses of quasi-elastic and resonant scattering~\cite{Andreopoulos:2009rq}, a correction of deep-inelastic scattering~\cite{Cooper-Sarkar:2011jtt}, and the ratio of the neutral- and charged-current cross sections.
The last two are the normalizations of the neutrino and muon events, $A_{\rm eff}$ and $s_\mu$.
Besides the LRI, we depart from the DeepCore analysis in four ways: our flux is that of MCEq instead of the Honda flux~\cite{Honda:2015fha} (Appendix~\ref{app:flux}); we include absorption and the spread of production heights (Sec.~\ref{sec:evolution}); and we let the four oscillation parameters that it fixes vary (Sec.~\ref{sec:statistics}).
Without an LRI, our fit reproduces its best fit (Appendix~\ref{app:fit_no_lri}).

\begin{figure}[t]
    \centering
    \includegraphics[width=\columnwidth]{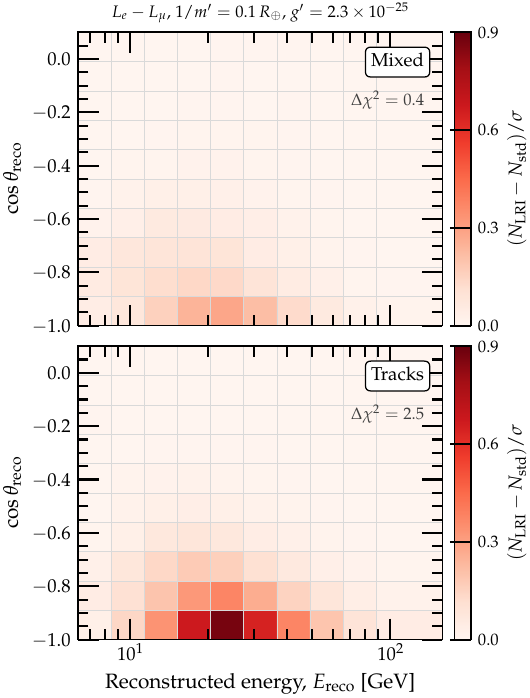}
    \caption{\textbf{\textit{Change in the expected events caused by a long-range interaction, in units of the uncertainty of each bin.}}
    For the benchmark interaction of Figs.~\ref{fig:oscillograms} and~\ref{fig:oscillation_probabilities}, with the best-fit oscillation parameters of the DeepCore analysis and nominal nuisance parameters.
    Here, $N_{\rm LRI}$ and $N_{\rm std}$ are the expected events with and without the interaction.
    The uncertainty $\sigma$ is that of Fig.~\ref{fig:data_fit}.
    The $\Delta\chi^2$ of each panel is the sum of the squares of its bins; freeing the nuisance parameters can only lower it.
    The change concentrates in steeply up-going tracks, where the interaction partly fills the oscillation minimum near 24~GeV (Fig.~\ref{fig:oscillation_probabilities}).
    See Sec.~\ref{sec:events} for details.}
    \label{fig:lri_bins}
\end{figure}

Figure~\ref{fig:lri_bins} shows how our benchmark LRI changes the expected events, bin by bin, with every parameter at its nominal value.
Almost every bin gains events, since the LRI fills part of the $\nu_\mu$ disappearance.
The change concentrates in the six bins of steeply up-going tracks at 15--37~GeV (Sec.~\ref{sec:sample}), where the filled minimum of Sec.~\ref{sec:evolution} lies; these bins give 68\% of $\Delta\chi^2$.
The larger changes of the probability below 10~GeV (\figu{oscillograms}) barely show, likely because their fast oscillations average out over the spread of true energies and directions within each bin.
The sensitivity of DeepCore to the LRI rests on a few bins of steeply up-going tracks.

\subsection{Statistical method}
\label{sec:statistics}

We adopt a frequentist approach.
For each hypothesis, we compare the expected and observed events with the test statistic of the DeepCore analysis~\cite{IceCubeCollaboration:2023wtb},
\begin{equation}
 \label{equ:chi2}
 \chi^2 = \sum_{\rm bins} \frac{(N - N_{\rm obs})^2}{N + (\sigma^{\rm sim})^2}
 + \sum_j \frac{(s_j - \hat s_j)^2}{\sigma_{s_j}^2} \;,
\end{equation}
where the first sum runs over the 200 bins, $N_{\rm obs}$ is the observed number of events, and $(\sigma^{\rm sim})^2$ is the variance of $N$ due to the finite size of the simulation.
The second sum adds a Gaussian prior, of mean $\hat s_j$ and width $\sigma_{s_j}$, for each nuisance parameter $s_j$ that has one.
We let $\theta_{23}$ and $\Delta m^2_{31}$ vary freely.
The other four oscillation parameters, which the DeepCore analysis fixes, vary with priors from the two-dimensional $\chi^2$ tables of the global fit NuFIT~6.1~\cite{Esteban:2024eli, NuFIT61} for the pairs ($\theta_{12}$, $\Delta m^2_{21}$) and ($\theta_{13}$, $\delta_{\rm CP}$).

At each mediator mass, we minimize $\chi^2$ over the oscillation and nuisance parameters, $\boldsymbol{\theta}$, at fixed $g'^2$, and compare it with its minimum over $g'^2$ as well,
\begin{multline}
 \label{equ:limit}
 \Delta\chi^2(g'^2; m') = \min_{\boldsymbol{\theta}}\, \chi^2(g'^2, m', \boldsymbol{\theta}) \\
 - \min_{g'^2, \boldsymbol{\theta}}\, \chi^2(g'^2, m', \boldsymbol{\theta}) \;.
\end{multline}
The sign of the potential is that of $k$ (Table~\ref{tab:symmetries}), which multiplies $g'^2$ in \equ{texture}.
For $L_\mu - L_\tau$, whose effective coupling through the mixing with the $Z$ may have either sign, we compute the limit for both signs.

The limit on $g'^2$ is one-sided, since $g'^2 \geq 0$.
When the true value lies at the boundary, $g'^2 = 0$, $\Delta\chi^2$ follows asymptotically an equal mixture of zero and a $\chi^2$ distribution with one degree of freedom~\cite{Cowan:2010js}, so the 90\%~C.L.\ upper limit lies at $\Delta\chi^2 = 1.64$, not at the value of 2.71 that Wilks' theorem~\cite{Wilks:1938dza} gives away from a boundary.
We use this asymptotic threshold rather than a Feldman--Cousins construction~\cite{Feldman:1997qc}, which the DeepCore analysis~\cite{IceCubeCollaboration:2023wtb} uses for its parameter intervals but which would require pseudo-experiments at every point of our scan.
Appendix~\ref{app:coverage} tests its coverage with pseudo-experiments: our 90\% and 95\%~C.L.\ limits cover the true coupling in 87\% and 94\% of them.
Thresholds that give exact coverage would weaken the limits by only about 2\% in $g'$.
To separate the sensitivity of the analysis from fluctuations of the data, we also compute expected limits, from the Asimov data set~\cite{Cowan:2010js}, in which the observed events equal those expected from our fit without an LRI.

The right panel of \figu{data_fit} shows the pulls of the observed events against our fit without an LRI.
The pull of a bin is $(N_{\rm obs} - N_{\rm fit})/\sigma$, where $N_{\rm fit}$ is the expected number of events at the fit and $\sigma$ is the square root of the denominator of the first sum of \equ{chi2}.
The fit of \figu{data_fit} fixes $\theta_{12}$, $\theta_{13}$, $\Delta m^2_{21}$, and $\delta_{\rm CP}$ at the values of the DeepCore analysis, so that we can compare it with that analysis (Appendix~\ref{app:fit_no_lri}).
It describes the data well, with $\chi^2 = 180.3$ for 200 bins and 20 free parameters; freeing these four parameters, as in the fits that set our limits, lowers $\chi^2$ by only 0.08.
The LRI of Sec.~\ref{sec:evolution} would add the most events in the six bins of steeply up-going tracks at 15--37~GeV, which carry most of our sensitivity (Sec.~\ref{sec:events}).
There, the data hold 425 events, 4\% fewer than the fit, a difference within their statistical uncertainty of 4.8\%.
This deficit is the likely reason why our observed limits are stronger than the expected ones (Sec.~\ref{sec:limits_lelmu}).

\section{Results}
\label{sec:results}

\begin{figure*}[t]
    \centering
    \includegraphics[width=\textwidth]{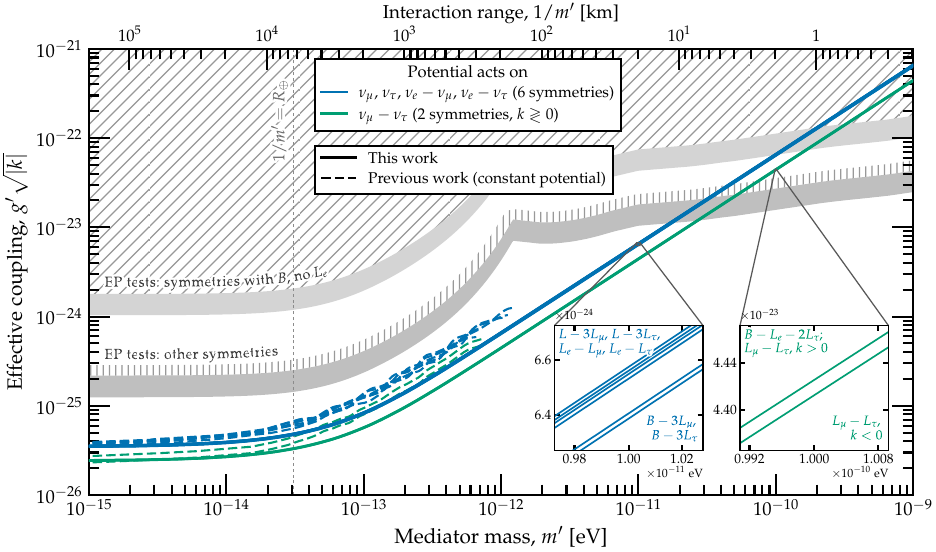}
    \caption{\textbf{\textit{Limits on the effective couplings of the symmetries that have an upper limit.}}
    The effective coupling $g'\sqrt{|k|}$ includes the charge factor $k$ of each symmetry (Table~\ref{tab:symmetries}).
    Each solid color shows several overlapping curves, one per symmetry, expanded in the insets.
    In these units, the limits of the six symmetries whose potential acts on $\nu_\mu$, $\nu_\tau$, $\nu_e - \nu_\mu$, or $\nu_e - \nu_\tau$ agree within 3\%.
    The six symmetries whose potential acts on $\nu_e$ alone have no upper limit (Sec.~\ref{sec:nue_only}); Appendix~\ref{app:more_limits} shows their excluded intervals, and these limits at 95\%~C.L.
    The dashed curves are the limits of \Ref~\cite{Garg:2026rmy}, read off its Fig.~1 and multiplied by $\sqrt{|k|}$; each ends where it leaves the axes of that figure.
    Its six curves of the first group differ by up to 8\% at $10^{-15}$~eV and by up to 35\% near $10^{-13}$~eV.
    In these units, each symmetry has its own bound from tests of the equivalence principle (EP), translated as in \figu{lri_limits}.
    Hatching marks the side they exclude.
    The symmetry $L_\mu - L_\tau$ has no such bound, since matter carries its charge only through the mixing of the $Z'$ with the $Z$.}
    \label{fig:all_limits}
\end{figure*}

\subsection{Limits on $L_e - L_\mu$}
\label{sec:limits_lelmu}

Figure~\ref{fig:lri_limits} shows our 90\%~C.L.\ upper limit on $g'$ for $L_e - L_\mu$.
Below about $10^{-14}$~eV, where the interaction range exceeds the Earth's radius, the whole Earth sources the potential; the limit barely changes with mass.
Around $1/m' \approx R_\oplus$, the limit bends upward, since matter farther than about one range from each path stops contributing.
Above $10^{-12}$~eV, a range of about 200~km, the potential follows the local density (Sec.~\ref{sec:potential}), so the limit grows in proportion to $m'$.
As for the benchmark (Sec.~\ref{sec:events}), the limit comes from the steeply up-going tracks near 24~GeV, where an LRI would add events.
In these tracks, the data hold 4\% fewer events than our fit without an LRI (Sec.~\ref{sec:statistics}); the Asimov data set, which equals that fit, has no such deficit.
This deficit likely makes the observed limit stronger than the expected one, by 13--15\% for every symmetry with an upper limit and every mass.

Our limit is stronger than the bounds from tests of the equivalence principle at ranges longer than about 12~km (Sec.~\ref{sec:bounds}).
It is also stronger, by 1.03--1.5 times, than the limit of \Ref~\cite{Garg:2026gwx} from the same data, wherever both exist.
(In its Fig.~4, that reference places its 90\%~C.L.\ limit where $\Delta\chi^2 = 2.71$, the threshold of our 95\%~C.L.\ limits, against which its limit is 0.93--1.4 times ours.)
The two analyses differ in more than the potential, however: \Ref~\cite{Garg:2026gwx} fixes $\theta_{12}$, $\theta_{13}$, $\Delta m^2_{21}$, and $\delta_{\rm CP}$, and uses the Honda flux.
Its bounds on the potential and on $g'$ imply a potential per unit $g'^2$ at the detector 1.5--3.2 times ours.

To isolate the effect of computing the potential along each path, we repeat our analysis with the potential held at its value at the detector along every path, everything else unchanged.
With this constant potential, our limit on $g'$ is weaker by a factor of 1.2 at $10^{-15}$~eV, of at least 2 for ranges shorter than about 600~km, and of 2.7 at $10^{-9}$~eV.
For the other seven symmetries, the factors differ from these by less than 6\%.
Inside the Earth, the potential exceeds its value at the detector over most of each path (\figu{vlri_trajectory}), so assuming a constant potential underestimates what the neutrinos feel.
The underestimate grows at shorter ranges, where the potential follows the density of the Earth more closely.
Up-going neutrinos cross the mantle and the core, which are denser than the crust and the water around the detector.

\subsection{All the symmetries}
\label{sec:limits_all}

Figure~\ref{fig:all_limits} shows the limits on the effective coupling $g'\sqrt{|k|}$ for the eight symmetries that have an upper limit.
In these units, the limits fall into two groups, within which they agree to about 3\%.
The two symmetries whose potential acts on $\nu_\mu - \nu_\tau$ have limits 1.45--1.47 times stronger than the other six.
The sensitivity rests on the difference between the potentials of $\nu_\mu$ and $\nu_\tau$ (Sec.~\ref{sec:nue_only}).
At equal $g'^2|k|$, this difference is twice as large for the texture $\nu_\mu - \nu_\tau$ as for the others, so the same effect needs half the value of $g'^2|k|$, that is, a limit on $g'\sqrt{|k|}$ lower by a factor of $\sqrt{2} \approx 1.41$.
Both signs of $k$ for $L_\mu - L_\tau$ give limits within 0.7\% of each other.

Figure~\ref{fig:all_limits} also shows the limits that \Ref~\cite{Garg:2026rmy} set on the same eight symmetries from the same data, with the potential held at its value at the detector.
Our limits are stronger, by 1.03--1.12 times at $10^{-15}$~eV and by about 1.5--2 times at ranges shorter than about 600~km.
Only its limit on $L_\mu - L_\tau$ is slightly stronger than ours, at the longest ranges.
In its Fig.~1, that limit coincides with the one on $B - L_e - 2L_\tau$, so we suspect that it uses a charge of 1 per neutron, not the $e/(\sin\theta_W \cos\theta_W)$ of its Eq.~(7).
With the charge of its Eq.~(7), it would lie 18\% higher in \figu{all_limits}, on that of $B - L_e - 2L_\tau$, above ours.

Besides the neutron charge for $L_\mu - L_\tau$, three effects make our limits differ from those of \Ref~\cite{Garg:2026rmy}.
First, computing the potential along each path makes our limits stronger, by a factor of 1.2 at $10^{-15}$~eV and of at least 2 at ranges shorter than about 600~km (Sec.~\ref{sec:limits_lelmu}).
Second, \Ref~\cite{Garg:2026rmy} places its 90\%~C.L.\ limits at $\Delta\chi^2 = 2.71$ (its Figs.~5--7), not 1.64, so they are about 10\% weaker.
Third, its potential per unit $g'^2$ at the detector is 1.4--1.6 times ours at $10^{-15}$~eV, so the same bound on the potential gives it a limit on $g'$ 1.2--1.3 times stronger.
To test whether other effects matter, we hold the potential at its value at the detector and place our limits at $\Delta\chi^2 = 2.71$, which removes the first two effects.
Our bounds on the LRI potential itself, before conversion into limits on $g'$, then lie within 4\% of those of \Refs~\cite{Garg:2026gwx, Garg:2026rmy}, for all eight symmetries.
So the other differences between the fits, among them the oscillation parameters that these references fix and their flux, together change the bounds by less than 4\%.
At $10^{-15}$~eV, the first and third nearly cancel, so the difference comes mostly from the second.

The six symmetries whose potential acts on $\nu_e$ alone have no upper limit (Sec.~\ref{sec:nue_only}).
Appendix~\ref{app:more_limits} shows the narrow intervals of $g'$ excluded for two of them, and the limits of \figu{all_limits} at 95\%~C.L.
We provide the limits on $g'$ at 90\% and 95\%~C.L.\ for every symmetry and mass, and the excluded intervals, as supplementary files~\cite{NuLRIEarthRepo}.

\subsection{Against other bounds}
\label{sec:bounds}

The $Z'$ that creates our potential also couples to ordinary matter, with the same coupling $g'$ and the charges of \equ{lagrangian}.
Between two bodies of matter, it adds to gravity a Yukawa force proportional to their charges; an atom with $Z$ protons and $N$ neutrons carries the charge $\tilde q = (a_e + a_p) Z + a_n N$~\cite{Wagner:2012ui, Coloma:2020gfv}.
Since the charge per unit mass differs between materials, bodies of different composition fall toward the Earth with different accelerations.
Tests of the equivalence principle bound this difference, and with it the strength of the force relative to gravity, $|\tilde\alpha| = g'^2/(4\pi G u^2)$, where $G$ is Newton's constant and $u$ is the atomic mass unit~\cite{Wagner:2012ui}.
This bounds $g'$ for every symmetry except $L_\mu - L_\tau$, whose charge ordinary matter does not carry; for it, these tests bound only the coupling that matter acquires through $Z$--$Z'$ mixing, not $G'$.

Below $m' \approx 10^{-12}$~eV, the strongest of these bounds comes from the MICROSCOPE satellite, which compares the free fall of titanium- and platinum-alloy test masses in orbit~\cite{MICROSCOPE:2022doy}.
For each symmetry, we compute the Yukawa field of our Earth model at its orbit.
Our bounds agree to within 6\% with those that \Ref~\cite{Fayet:2018cjy} derived from the first MICROSCOPE result, once rescaled to the final one.

Above $10^{-12}$~eV, the strongest bound comes from the E\"ot-Wash rotating torsion balance, which compares test bodies of beryllium with ones of titanium and of aluminum~\cite{Wagner:2012ui}.
Reference~\cite{Wagner:2012ui} gives this bound versus the range only for $\tilde q = B - L$.
As \Ref~\cite{Coloma:2020gfv} did for an earlier E\"ot-Wash result~\cite{Schlamminger:2007ht}, we rescale it to each symmetry: the signal of each pair of test bodies is proportional to $\tilde\alpha$, to the difference between their charges per unit mass, and to that of the source, for which we take rock.
For ranges much longer than the Earth's radius, this reproduces the bounds of \Ref~\cite{Wagner:2012ui} for the charges of the symmetries in our figures to within 8\%.
(We convert both bounds to 90\%~C.L., the level of our limits, by placing each 1.645 standard deviations above the magnitude of the measured signal, which for E\"ot-Wash we take from Table~3 of \Ref~\cite{Wagner:2012ui}.)
Tests of the inverse-square law of gravity give weaker bounds throughout our window~\cite{Adelberger:2003zx}.

Figures~\ref{fig:lri_limits} and \ref{fig:all_limits} compare our limits with these bounds.
For each symmetry that the tests constrain, our limit is the stronger one below a crossing mass and the weaker one above it.
Below $10^{-12}$~eV, our limits are stronger by factors of 2.8--11, and by 41--87 for $B - 3L_\mu$ and $B - 3L_\tau$.
Above $10^{-12}$~eV, the potential approaches the contact limit, $\Psi_f \to n_f/m'^2$ (Sec.~\ref{sec:potential}), so our limits grow in proportion to $m'$: by a factor of 1000 up to $10^{-9}$~eV, against 3--6 for the bounds from the tests.
The two cross at $m' = 1.7 \times 10^{-11}$--$2.1 \times 10^{-10}$~eV, ranges of 1--12~km.

The tests are weakest for the symmetries with $B$ and no $L_e$, such as $B - 3L_\mu$ and $B - 3L_\tau$, since matter carries their charge only through its baryons.
The number of baryons per unit mass is nearly the same in every material, so this force barely changes the relative acceleration of test bodies.
The potential that neutrinos feel, instead, grows with the total charge of the Earth.
Among the symmetries without an upper limit, the excluded intervals of $B_y + L_\mu + L_\tau$ lie below its bound from these tests, excluding couplings that the tests allow (Appendix~\ref{app:more_limits}).

\section{Summary and outlook}
\label{sec:summary}

A new long-range force between neutrinos and matter, different for each flavor, would let the electrons and nucleons of the Earth act on the atmospheric neutrinos that cross it.
Together, they would create a potential that changes how these neutrinos oscillate.
Inside the Earth, this potential changes along the path of each neutrino, more so the shorter the range of the force.
We computed the potential at every point and fit the public IceCube DeepCore sample with the same nuisance parameters as IceCube's own analysis of it.
Unlike that analysis, we also fit the oscillation parameters that it held fixed.

\textbf{At ranges longer than about 1--12~km, our limits on such forces are stronger than those from tests of the equivalence principle, by up to almost two orders of magnitude.}
\textbf{They are also stronger than the limits obtained by holding the potential at its value at the detector, as earlier analyses did.}
The difference grows from about 20\% at the longest ranges to about a factor of 3 at the shortest.
We set these limits for the forces of a broad set of new $U(1)'$ gauge symmetries.
Those whose potential differs between $\nu_\mu$ and $\nu_\tau$ have upper limits.
Once their charges are accounted for, the strongest limits are for those whose potential separates $\nu_\mu$ from $\nu_\tau$ the most.
Those whose potential is the same for $\nu_\mu$ and $\nu_\tau$ have none: such a potential suppresses the mixing of $\nu_e$ but leaves unchanged the minimum of the $\nu_\mu$ survival probability, on which our sensitivity rests.

Detectors under construction or planned, such as the IceCube Upgrade~\cite{Ishihara:2019aao}, KM3NeT/ORCA~\cite{KM3Net:2016zxf}, Hyper-Kamiokande~\cite{Hyper-Kamiokande:2018ofw}, and DUNE~\cite{DUNE:2020ypp}, could tighten these limits.
For every range that we study, the potential inside the Earth varies along each path, by a factor of about 2 or more between the detector and the center.
So analyses of their data should, like ours, follow the potential along the path of each neutrino.

\begin{acknowledgments}
DFC acknowledges the support of the Vicerrectorado de Investigaci\'on PUCP, of the Facultad de Ciencias e Ingenier\'ia PUCP via the Fondo Extraordinario de Apoyo a la Investigaci\'on para Estudiantes (FEAPE), and of the Research Experience for Peruvian Undergraduates (REPU) program, and thanks the Niels Bohr Institute for hospitality during the completion of this work.
MB is supported by {\sc Villum Fonden} under project no.~29388. This work used the Tycho supercomputer hosted at the SCIENCE High Performance Computing Center at the University of Copenhagen. During the preparation of this work, the authors used Claude for research assistance in code development, numerical auditing, figure preparation, and editorial revision. All scientific judgments, validation decisions, and responsibility for the manuscript remain with the authors.
\end{acknowledgments}

\appendix
\counterwithin{figure}{section}
\counterwithin{table}{section}

\section{The potential of a layered Earth}
\label{app:lri_potential}

In a spherical Earth, the density depends only on the distance from the center, so the angular part of \equ{psi} can be integrated over each spherical shell of matter.
A shell of radius $r'$ contributes to $\Psi_f$ at radius $r$ in proportion to $e^{-m'|r - r'|} - e^{-m'(r + r')}$, which gives
\begin{multline}
 \label{equ:radial}
 \Psi_f(r) = \frac{1}{2m'r} \int_0^{R_\oplus} dr'\, r'\, n_f(r') \\
 \times \left[ e^{-m'|r - r'|} - e^{-m'(r + r')} \right] \;.
\end{multline}

In a layer $j$ of constant density $n_j$, between radii $r_{j-1}$ and $r_j$, the integral of \equ{radial} has a closed form.
The matter of the layer above the neutrino, from $u = \max(r_{j-1}, r)$ to $r_j$, contributes
\begin{multline}
 \label{equ:closed_above}
 \frac{n_j \left(1 - e^{-2m'r}\right)}{2m'r}
 \bigg[ e^{-m'(u - r)} \left(\frac{u}{m'} + \frac{1}{m'^2}\right) \\
 - e^{-m'(r_j - r)} \left(\frac{r_j}{m'} + \frac{1}{m'^2}\right) \bigg] \;,
\end{multline}
and the matter below it, from $r_{j-1}$ to $v = \min(r_j, r)$, contributes
\begin{multline}
 \label{equ:closed_below}
 \frac{n_j}{2m'r} \bigg[ e^{-m'(r - x)} \left(\frac{x}{m'} - \frac{1}{m'^2}\right) \\
 + e^{-m'(r + x)} \left(\frac{x}{m'} + \frac{1}{m'^2}\right) \bigg]_{x = r_{j-1}}^{x = v} \;.
\end{multline}
Summing these over the 12 layers gives $\Psi_f$ at any radius, inside or outside the Earth.

The closed form shows how the potential follows the density.
Deep inside a layer, at distances from its boundaries much larger than $1/m'$, $\Psi_f$ tends to $n_j/m'^2$, the contact limit.
At a boundary between two layers thicker than $1/m'$, it tends to the average of their two contact values.
It is continuous across every boundary: the potential smooths each jump of the density over about one interaction range.
This is why it follows the density only for ranges shorter than the layers (Sec.~\ref{sec:potential}).

\section{Atmospheric neutrino flux}
\label{app:flux}

\begin{figure}[t]
    \centering
    \includegraphics[width=\columnwidth]{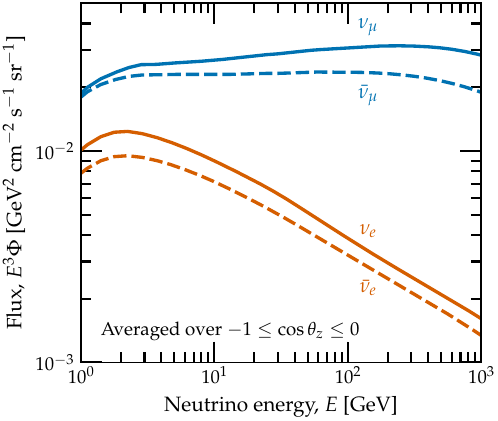}
    \caption{\textbf{\textit{Atmospheric neutrino flux used in this work.}}
    The flux is computed with MCEq~\cite{Fedynitch:2015zma} and averaged over the twelve months.
    It does not include oscillations or absorption inside the Earth.
    Muon neutrinos and antineutrinos dominate, increasingly so at higher energies, where more atmospheric muons, the main source of electron neutrinos, reach the ground before decaying.
    See Appendix~\ref{app:flux} for details.}
    \label{fig:flux}
\end{figure}

Figure~\ref{fig:flux} shows the atmospheric neutrino flux of our fit, before the neutrinos cross the Earth.
It is the flux $\Phi_\alpha$ of \equ{events}, with the flux nuisance parameters at their nominal values.
The DeepCore analysis uses the Honda flux~\cite{Honda:2015fha}, but it computes the uncertainties of that flux with MCEq~\cite{IceCubeCollaboration:2023wtb}.
For each of its six parameters of hadron production, it computes the change of the flux with the SIBYLL-2.3c model of hadronic interactions and the GSF model of cosmic rays~\cite{Dembinski:2017zsh}, and finds that the choice of these models has a negligible effect.
With MCEq, the flux and its uncertainties come from one calculation.

We use SIBYLL-2.3d~\cite{Riehn:2019jet}, the successor of SIBYLL-2.3c, and version 2026.1 of the GSF model~\cite{Fedynitch:2026ugq}.
We average the modulation of the cosmic rays by the Sun over the years of the data, 2011--2019.
We take the density of the air from the NRLMSISE-00 model of the atmosphere~\cite{Picone:2002}, above the site where each path enters the atmosphere, for up-going and down-going directions alike.
At the South Pole, the air ends at the surface of the ice, 2.8~km above sea level.

The atmosphere changes with the season, so we compute the flux for each month and use the mean of the twelve.
Since the expected events are linear in the flux, this mean gives the same events as a calculation month by month.
(Weighting the months by their livetime instead would shift the flux by about 0.1\%.)
We keep two parts of the MCEq flux that the Honda flux leaves out, both small below 100~GeV: the neutrinos from the decays of charmed hadrons, at most 0.7\% of the $\nu_e$ flux, and the flux of $\nu_\tau$, less than $10^{-4}$ of that of $\nu_\mu$.

Of the seven flux parameters of the analysis, the change of the spectral index multiplies the flux by the same factor at every height.
IceCube gives the changes due to the six parameters of hadron production only for the total flux, not by height, so we give them the distribution in height of our flux.
At one width of its prior, each of these changes is at most 10\% of the flux between 3 and 100~GeV.
Giving the whole flux a single production height alters the expected events by 0.5\% at most (Sec.~\ref{sec:evolution}), so we estimate that our choice for these changes shifts the expected events by about 0.05\%.

Our flux differs from the Honda flux of the DeepCore analysis mostly by an overall factor, which the normalization $A_{\rm eff}$ absorbs.
At 10 and 50~GeV, over all directions, it is 0.81--0.97 times the Honda flux.
With the oscillation parameters of the DeepCore best fit and every nuisance parameter at its nominal value, it gives 0.87 times the neutrino events of the Honda flux.
Normalized to the same total, the two fluxes give events that differ by $-3\%$ to $+7\%$ per bin, most of all in the highest bins of reconstructed energy, where our flux gives more events.

The flux parameters of the analysis absorb this difference.
Fitting the events of the Honda flux to those of ours, with only the normalization and the seven flux parameters free, within the priors of the analysis, leaves the two predictions within 0.43 statistical errors of each other in every bin.
The fit moves the spectral index and the six parameters of hadron production by at most 0.35 widths of their priors.
Within the statistical errors of the sample, our flux is the Honda flux with its flux parameters moved well within their priors.

\section{Parameters of the fit}
\label{app:parameters}

\begin{table*}[t]
 \caption{\label{tab:parameters}\textbf{\textit{Parameters of the fit, with their priors and their values at the fit without a long-range interaction.}}
 The groups are those of Table~IV of \Ref~\cite{IceCubeCollaboration:2023wtb}, each with its baseline.
 A prior is a Gaussian of the width given, except for the two pairs of oscillation parameters, whose prior is a two-dimensional $\Delta\chi^2$ table of NuFIT~6.1; their nominal values are the minima of these tables.
 The pull is the best fit minus the nominal value, in units of the prior width, so that its square is the term that the prior adds to $\chi^2$; for each pair with a NuFIT prior, it is the square root of the $\Delta\chi^2$ of its table at the best fit.
 The axial masses are in units of their prior widths, $+25\%$ and $-15\%$ of 0.99~GeV (CCQE) and $\pm 20\%$ of 1.12~GeV (CCRES); the changes of the hadron yields are fractions of the nominal yields.
 See Appendix~\ref{app:parameters} for details.}
 \begin{ruledtabular}
 \begin{tabular}{l c c c c c}
 Parameter & Nominal & Prior & Range & Best fit & Pull \\
 \hline
 \multicolumn{6}{l}{\textit{Long-range interaction}} \\
 Coupling, $g'^2$ & 0 & None & $\geq 0$ & --- & --- \\
 Mediator mass, $m'$ [eV] & --- & --- & 25 values, $10^{-15}$--$10^{-9}$ & --- & --- \\
 \multicolumn{6}{l}{\textit{Oscillations, baseline from NuFIT~6.1~\cite{Esteban:2024eli, NuFIT61}, normal ordering}} \\
 $\sin^2\theta_{23}$ & --- & None & $[0.3, 0.7]$ & 0.504 & --- \\
 $\Delta m^2_{31}$ [$10^{-3}$~eV$^2$] & --- & None & $[1.5, 3.5]$ & 2.51 & --- \\
 $\sin^2\theta_{12}$ & 0.308 & \multirow{2}{*}{NuFIT~6.1 table} & $[0.27, 0.35]$ & 0.309 & \multirow{2}{*}{0.00} \\
 $\Delta m^2_{21}$ [$10^{-5}$~eV$^2$] & 7.53 & & $[6.8, 8.3]$ & 7.53 & \\
 $\sin^2\theta_{13}$ & 0.0225 & \multirow{2}{*}{NuFIT~6.1 table} & $[0.019, 0.026]$ & 0.0225 & \multirow{2}{*}{0.00} \\
 $\delta_{\rm CP}$ [$^\circ$] & 210 & & $[0, 360]$ & 212 & \\
 \multicolumn{6}{l}{\textit{Detector, baseline from calibration data}} \\
 Efficiency of the optical modules & 1.00 & 0.10 & $[0.5, 1.5]$ & 1.047 & 0.47 \\
 Hole ice, $p_0$ & 0.10 & None & $[-2, 1]$ & $-0.268$ & --- \\
 Hole ice, $p_1$ & $-0.05$ & None & $[-0.2, 0.2]$ & $-0.040$ & --- \\
 Bulk ice, absorption & 1.00 & None & $[0.90, 1.10]$ & 0.977 & --- \\
 Bulk ice, scattering & 1.00 & None & $[0.95, 1.15]$ & 0.993 & --- \\
 \multicolumn{6}{l}{\textit{Flux, baseline from MCEq (DeepCore analysis: Honda flux)}} \\
 Spectral index, $\Delta\gamma$ & 0 & 0.10 & $[-0.5, 0.5]$ & 0.036 & 0.36 \\
 $\pi^\pm$ yields, A--F & 0 & 0.30 & $[-1.5, 1.5]$ & 0.008 & 0.03 \\
 $\pi^\pm$ yields, G & 0 & 0.30 & $[-1.5, 1.5]$ & $-0.034$ & $-0.11$ \\
 $\pi^\pm$ yields, H & 0 & 0.15 & $[-0.75, 0.75]$ & $-0.006$ & $-0.04$ \\
 $K^+$ yields, W & 0 & 0.40 & $[-2, 2]$ & 0.022 & 0.05 \\
 $K^-$ yields, W & 0 & 0.40 & $[-2, 2]$ & $-0.025$ & $-0.06$ \\
 $K^+$ yields, Y & 0 & 0.30 & $[-1.5, 1.5]$ & 0.023 & 0.08 \\
 \multicolumn{6}{l}{\textit{Cross sections, baseline from GENIE}} \\
 $M_A^{\rm CCQE}$ & 0 & 1 & $[-2, 2]$ & $-0.14$ & $-0.14$ \\
 $M_A^{\rm CCRES}$ & 0 & 1 & $[-2, 2]$ & 0.25 & 0.25 \\
 $\sigma_{\rm NC}/\sigma_{\rm CC}$ & 1.00 & 0.20 & $[0, 2]$ & 1.103 & 0.51 \\
 DIS, from GENIE (0) to CSMS (1) & 0 & 1.0 & $[-5, 5]$ & 0.20 & 0.20 \\
 \multicolumn{6}{l}{\textit{Atmospheric muons, template of the data release}} \\
 Normalization, $s_\mu$ & 1 & None & $[0, 3]$ & 1.66 & --- \\
 \multicolumn{6}{l}{\textit{Normalization of the neutrino events}} \\
 $A_{\rm eff}$ & 1 & None & $[0, 3]$ & 0.978 & --- \\
 \end{tabular}
 \end{ruledtabular}
\end{table*}

Table~\ref{tab:parameters} lists the parameters of our fit, in the groups of Table~IV of the DeepCore analysis~\cite{IceCubeCollaboration:2023wtb}.
The nuisance parameters, their nominal values, and their priors are those of that analysis; only the baseline of the flux differs (Appendix~\ref{app:flux}).
The DeepCore analysis fixes the oscillation parameters other than $\theta_{23}$ and $\Delta m^2_{31}$; we let them vary, with the priors of Sec.~\ref{sec:statistics}.
The ranges are our choice: the bounds of the DeepCore analysis for the ice, two prior widths for the axial masses, five widths for the other parameters with a Gaussian prior, $[0, 3]$ for the two normalizations, and, for $\Delta m^2_{31}$, the range over which the data release gives the hypersurfaces.
For the parameters with NuFIT priors, the range covers every value at which the NuFIT table, profiled over the other parameter of its pair, is below $\Delta\chi^2 = 25$, as a Gaussian prior is within five widths; the range of $\delta_{\rm CP}$ is the full circle.
The best-fit values and pulls are those of our fit without an LRI, in which the oscillation and nuisance parameters vary as they do in the scan.

\section{Fit without a long-range interaction}
\label{app:fit_no_lri}

\begin{figure}[t]
    \centering
    \includegraphics[width=0.98\columnwidth]{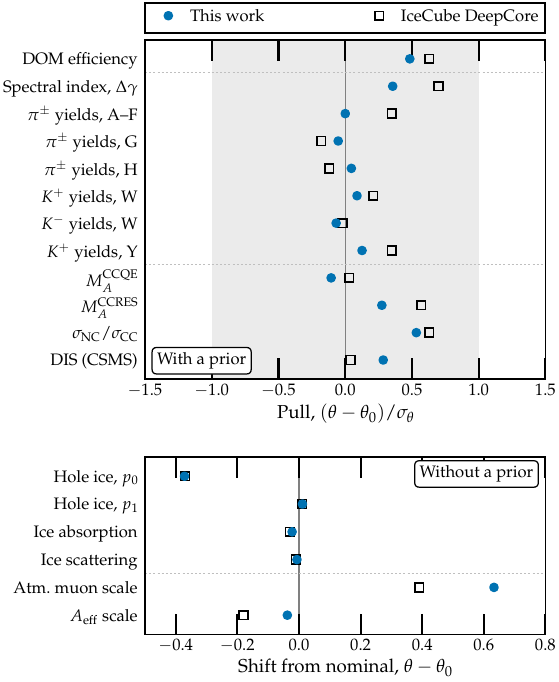}
    \caption{\textbf{\textit{Nuisance parameters at the fit without a long-range interaction.}}
    Here, $\theta_0$ is the nominal value of a parameter and $\sigma_\theta$ the width of its Gaussian prior; the gray band spans one width.
    Dotted lines separate the parameters of the detector, the flux, and the cross sections (top), and those of the detector from the normalizations (bottom).
    The DeepCore values are those of \Ref~\cite{IceCubeCollaboration:2023wtb}, with the Honda flux: the pulls of its Table~IV for the parameters with a prior, and its fit values for the others.
    Both fits keep every parameter with a prior within one width of it.
    See Appendix~\ref{app:fit_no_lri} for details.}
    \label{fig:pulls}
\end{figure}

Figure~\ref{fig:pulls} shows the nuisance parameters at our fit to the data without an LRI, next to those of the DeepCore analysis~\cite{IceCubeCollaboration:2023wtb}.
This is the fit of \figu{data_fit}.
Like the DeepCore analysis, it keeps the oscillation parameters other than $\theta_{23}$ and $\Delta m^2_{31}$ fixed at the values of that analysis, so that the two fits can be compared directly.
We find $\sin^2\theta_{23} = 0.510$ and $\Delta m^2_{32} = 2.44 \times 10^{-3}$~eV$^2$, against $0.51 \pm 0.05$ and $(2.41 \pm 0.07) \times 10^{-3}$~eV$^2$ in the DeepCore analysis, differences of less than 0.1 and of 0.4 times its uncertainties.
Our best fit predicts 22,016 events, against 22,012 in the DeepCore fit and 21,914 observed.
The charged-current events of $\nu_\mu$ and of $\nu_\tau$, and the other neutrino events, each differ from those of the DeepCore analysis by at most 2.3\%.
The $\Delta\chi^2$ of our limits is measured instead from the fit with every oscillation parameter free and the priors of Sec.~\ref{sec:statistics}, whose values Table~\ref{tab:parameters} lists.

The detector parameters nearly coincide with those of the DeepCore analysis.
The parameters of the flux lie closer to their nominal values: the data need smaller changes of the MCEq flux than of the Honda flux.
Our normalization of the neutrino events, $A_{\rm eff}$, exceeds that of the DeepCore analysis by nearly the factor by which the MCEq flux falls short of the Honda flux in events (Appendix~\ref{app:flux}).
The atmospheric muons, whose normalization $s_\mu$ has no prior, account for 131 more events than in the DeepCore analysis, 0.6\% of the sample.
Together, the MCEq flux, the absorption in the Earth, and the folding over production heights shift each parameter with a prior by at most 0.35 widths of that prior.

\section{Coverage of the limits}
\label{app:coverage}

Table~\ref{tab:coverage} lists the coverage of our upper limits, which we test with pseudo-experiments for $L_e - L_\mu$ at three masses: $10^{-15}$~eV, on the plateau of the limit; $3.2 \times 10^{-13}$~eV, where it bends upward; and $3.2 \times 10^{-12}$~eV, where it grows in proportion to $m'$.
For each confidence level and mass, we generate 200 pseudo-experiments, with the true coupling at our expected limit at that level and mass (Sec.~\ref{sec:statistics}).
Each draws the events of every bin from a Poisson distribution around the expected events of our fit to the data without an LRI, with the LRI at the true coupling added.
Our fit to each pseudo-experiment is the same as to the data, with every oscillation and nuisance parameter free.
The priors keep their nominal centers: the pseudo-experiments vary the events, not the measurements behind the priors.
An upper limit covers the true coupling when $\Delta\chi^2$ at the true coupling is at most the threshold, or when the best fit lies above the true coupling, since the limit then lies above it too.

\begin{table}[t]
 \caption{\label{tab:coverage}\textbf{\textit{Coverage of the upper limits on $L_e - L_\mu$.}}
 The fraction of 200 pseudo-experiments whose limit lies above the true coupling, for each mass and confidence level, with its binomial uncertainty.
 The true coupling is our expected limit at that level and mass.
 See Appendix~\ref{app:coverage} for details.}
 \begin{ruledtabular}
 \begin{tabular}{l c c}
 $m'$ [eV] & 90\%~C.L. & 95\%~C.L. \\
 \hline
 $10^{-15}$ & $0.870 \pm 0.024$ & $0.925 \pm 0.019$ \\
 $3.2 \times 10^{-13}$ & $0.865 \pm 0.024$ & $0.955 \pm 0.015$ \\
 $3.2 \times 10^{-12}$ & $0.875 \pm 0.023$ & $0.935 \pm 0.017$ \\
 \end{tabular}
 \end{ruledtabular}
\end{table}

Over the three masses together, the 90\%~C.L.\ limits cover the true coupling in a fraction $0.870 \pm 0.014$ of the pseudo-experiments, 2.2 standard deviations below the nominal 0.90.
The 95\%~C.L.\ limits cover it in $0.938 \pm 0.010$, 1.2 standard deviations below the nominal 0.95.
The thresholds that would give the nominal coverage are 1.85 and 3.02, instead of 1.64 and 2.71; with them, our limits on $L_e - L_\mu$ would be 1.4--2.3\% weaker in $g'$.
We keep the asymptotic thresholds of Sec.~\ref{sec:statistics}.
In 6--18 of the 200 pseudo-experiments of each set, a fit ended at the edge of a parameter range, or the fit in the other octant of $\theta_{23}$ did not finish; leaving them out changes the coverage by at most 0.007.

\section{Limits at 95\%~C.L.\ and excluded intervals}
\label{app:more_limits}

\begin{figure*}[t]
    \centering
    \includegraphics[width=\textwidth]{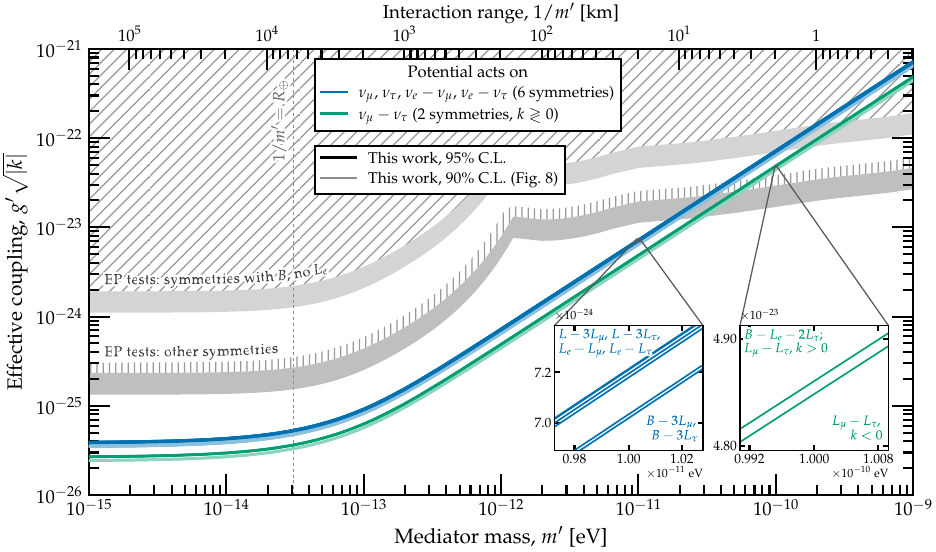}
    \caption{\textbf{\textit{Limits on the effective couplings at 95\%~C.L.}}
    Same as \figu{all_limits}, at 95\%~C.L., with the bounds from tests of the equivalence principle at 95\%~C.L.\ too and without the limits of \Ref~\cite{Garg:2026rmy}, which are at 90\%~C.L.
    The six symmetries whose potential acts on $\nu_e$ alone have no upper limit and no excluded interval at this level.
    See Appendix~\ref{app:more_limits} for details.}
    \label{fig:all_limits_cl95}
\end{figure*}
\begin{figure*}[t]
    \centering
    \includegraphics[width=\textwidth]{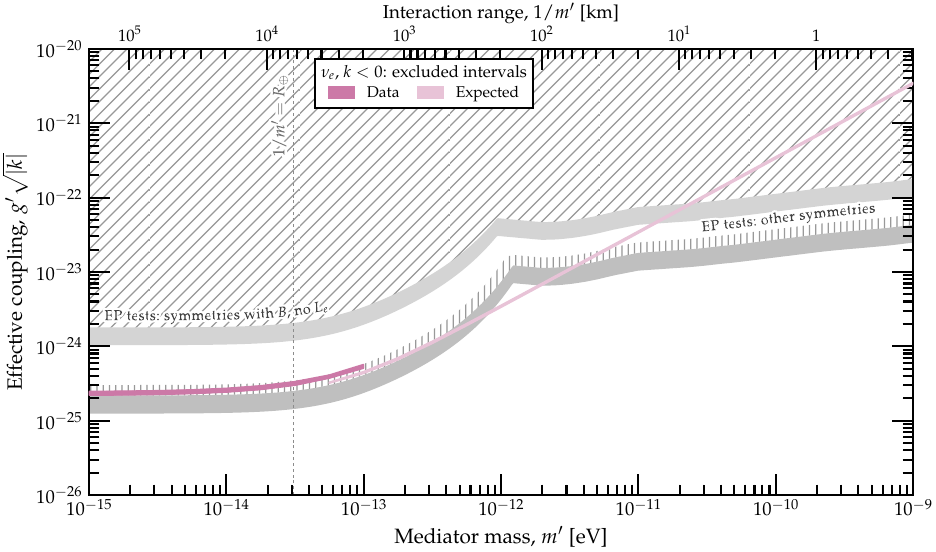}
    \caption{\textbf{\textit{Bounded intervals of couplings excluded at 90\%~C.L.\ for the two symmetries whose potential acts on $\nu_e$ alone with $k < 0$.}}
    At each mass, the couplings between the lower and the upper edge of a band are excluded.
    The two symmetries agree within 3\% in these units, so their bands overlap.
    The expected intervals are those of the Asimov data set (Sec.~\ref{sec:statistics}).
    The gray bands are the bounds from tests of the equivalence principle, as in \figu{all_limits}: the bound on $B_y + L_\mu + L_\tau$ lies within the upper band; the bound on $L_e + 2L_\mu + 2L_\tau$ lies within the lower one.
    See Appendix~\ref{app:more_limits} for details.}
    \label{fig:intervals}
\end{figure*}

\smallskip
\textit{Limits at 95\%~C.L.}---Figure~\ref{fig:all_limits_cl95} shows the limits of \figu{all_limits} at 95\%~C.L., where the threshold on $\Delta\chi^2$ is 2.71 instead of 1.64 (Sec.~\ref{sec:statistics}).
The same eight symmetries have upper limits, about 10\% weaker than at 90\%~C.L.
The bounds from tests of the equivalence principle are drawn at 95\%~C.L.\ as well.

\smallskip
\textit{Excluded intervals.}---Figure~\ref{fig:intervals} shows the bounded intervals of couplings that the data exclude at 90\%~C.L.\ for $L_e + 2L_\mu + 2L_\tau$ and $B_y + L_\mu + L_\tau$, the two symmetries whose potential acts on $\nu_e$ alone with $k < 0$ (Sec.~\ref{sec:nue_only}).
In each interval, $\Delta\chi^2$ rises just above the 90\%~C.L.\ threshold, to at most 2.35, so no interval survives at 95\%~C.L.
On the Asimov data set, intervals appear only from $5.6 \times 10^{-14}$~eV up; at lower masses, the expected $\Delta\chi^2$ stays below the threshold.
So, below that mass, the intervals of the data stem from an upward fluctuation.
For $B_y + L_\mu + L_\tau$, the intervals lie below the bound from tests of the equivalence principle, so they exclude couplings that those tests allow.
For $L_e + 2L_\mu + 2L_\tau$, those tests already exclude them.
Our pseudo-experiments test the coverage of upper limits (Appendix~\ref{app:coverage}), not that of these intervals.


\begin{thebibliography}{75}%
\makeatletter
\providecommand \@ifxundefined [1]{%
 \@ifx{#1\undefined}
}%
\providecommand \@ifnum [1]{%
 \ifnum #1\expandafter \@firstoftwo
 \else \expandafter \@secondoftwo
 \fi
}%
\providecommand \@ifx [1]{%
 \ifx #1\expandafter \@firstoftwo
 \else \expandafter \@secondoftwo
 \fi
}%
\providecommand \natexlab [1]{#1}%
\providecommand \enquote  [1]{``#1''}%
\providecommand \bibnamefont  [1]{#1}%
\providecommand \bibfnamefont [1]{#1}%
\providecommand \citenamefont [1]{#1}%
\providecommand \href@noop [0]{\@secondoftwo}%
\providecommand \href [0]{\begingroup \@sanitize@url \@href}%
\providecommand \@href[1]{\@@startlink{#1}\@@href}%
\providecommand \@@href[1]{\endgroup#1\@@endlink}%
\providecommand \@sanitize@url [0]{\catcode `\\12\catcode `\$12\catcode `\&12\catcode `\#12\catcode `\^12\catcode `\_12\catcode `\%12\relax}%
\providecommand \@@startlink[1]{}%
\providecommand \@@endlink[0]{}%
\providecommand \url  [0]{\begingroup\@sanitize@url \@url }%
\providecommand \@url [1]{\endgroup\@href {#1}{\urlprefix }}%
\providecommand \urlprefix  [0]{URL }%
\providecommand \Eprint [0]{\href }%
\providecommand \doibase [0]{https://doi.org/}%
\providecommand \selectlanguage [0]{\@gobble}%
\providecommand \bibinfo  [0]{\@secondoftwo}%
\providecommand \bibfield  [0]{\@secondoftwo}%
\providecommand \translation [1]{[#1]}%
\providecommand \BibitemOpen [0]{}%
\providecommand \bibitemStop [0]{}%
\providecommand \bibitemNoStop [0]{.\EOS\space}%
\providecommand \EOS [0]{\spacefactor3000\relax}%
\providecommand \BibitemShut  [1]{\csname bibitem#1\endcsname}%
\let\auto@bib@innerbib\@empty
\bibitem [{\citenamefont {Lee}\ and\ \citenamefont {Yang}(1955)}]{Lee:1955vk}%
  \BibitemOpen
  \bibfield  {author} {\bibinfo {author} {\bibfnamefont {T.~D.}\ \bibnamefont {Lee}}\ and\ \bibinfo {author} {\bibfnamefont {C.-N.}\ \bibnamefont {Yang}},\ }\bibfield  {title} {\bibinfo {title} {{Conservation of Heavy Particles and Generalized Gauge Transformations}},\ }\href {https://doi.org/10.1103/PhysRev.98.1501} {\bibfield  {journal} {\bibinfo  {journal} {Phys. Rev.}\ }\textbf {\bibinfo {volume} {98}},\ \bibinfo {pages} {1501} (\bibinfo {year} {1955})}\BibitemShut {NoStop}%
\bibitem [{\citenamefont {Okun}(1996)}]{Okun:1995dn}%
  \BibitemOpen
  \bibfield  {author} {\bibinfo {author} {\bibfnamefont {L.}~\bibnamefont {Okun}},\ }\bibfield  {title} {\bibinfo {title} {{Leptons and photons}},\ }\href {https://doi.org/10.1016/0370-2693(96)00586-2} {\bibfield  {journal} {\bibinfo  {journal} {Phys. Lett. B}\ }\textbf {\bibinfo {volume} {382}},\ \bibinfo {pages} {389} (\bibinfo {year} {1996})},\ \Eprint {https://arxiv.org/abs/hep-ph/9512436} {arXiv:hep-ph/9512436} \BibitemShut {NoStop}%
\bibitem [{\citenamefont {He}\ \emph {et~al.}(1991{\natexlab{a}})\citenamefont {He}, \citenamefont {Joshi}, \citenamefont {Lew},\ and\ \citenamefont {Volkas}}]{He:1990pn}%
  \BibitemOpen
  \bibfield  {author} {\bibinfo {author} {\bibfnamefont {X.~G.}\ \bibnamefont {He}}, \bibinfo {author} {\bibfnamefont {G.~C.}\ \bibnamefont {Joshi}}, \bibinfo {author} {\bibfnamefont {H.}~\bibnamefont {Lew}},\ and\ \bibinfo {author} {\bibfnamefont {R.~R.}\ \bibnamefont {Volkas}},\ }\bibfield  {title} {\bibinfo {title} {{New $Z'$ phenomenology}},\ }\href {https://doi.org/10.1103/PhysRevD.43.R22} {\bibfield  {journal} {\bibinfo  {journal} {Phys. Rev. D}\ }\textbf {\bibinfo {volume} {43}},\ \bibinfo {pages} {22} (\bibinfo {year} {1991}{\natexlab{a}})}\BibitemShut {NoStop}%
\bibitem [{\citenamefont {He}\ \emph {et~al.}(1991{\natexlab{b}})\citenamefont {He}, \citenamefont {Joshi}, \citenamefont {Lew},\ and\ \citenamefont {Volkas}}]{He:1991qd}%
  \BibitemOpen
  \bibfield  {author} {\bibinfo {author} {\bibfnamefont {X.-G.}\ \bibnamefont {He}}, \bibinfo {author} {\bibfnamefont {G.~C.}\ \bibnamefont {Joshi}}, \bibinfo {author} {\bibfnamefont {H.}~\bibnamefont {Lew}},\ and\ \bibinfo {author} {\bibfnamefont {R.~R.}\ \bibnamefont {Volkas}},\ }\bibfield  {title} {\bibinfo {title} {{Simplest $Z'$ model}},\ }\href {https://doi.org/10.1103/PhysRevD.44.2118} {\bibfield  {journal} {\bibinfo  {journal} {Phys. Rev. D}\ }\textbf {\bibinfo {volume} {44}},\ \bibinfo {pages} {2118} (\bibinfo {year} {1991}{\natexlab{b}})}\BibitemShut {NoStop}%
\bibitem [{\citenamefont {Foot}(1991)}]{Foot:1990mn}%
  \BibitemOpen
  \bibfield  {author} {\bibinfo {author} {\bibfnamefont {R.}~\bibnamefont {Foot}},\ }\bibfield  {title} {\bibinfo {title} {{New Physics From Electric Charge Quantization?}},\ }\href {https://doi.org/10.1142/S0217732391000543} {\bibfield  {journal} {\bibinfo  {journal} {Mod. Phys. Lett. A}\ }\textbf {\bibinfo {volume} {6}},\ \bibinfo {pages} {527} (\bibinfo {year} {1991})}\BibitemShut {NoStop}%
\bibitem [{\citenamefont {Foot}\ \emph {et~al.}(1990)\citenamefont {Foot}, \citenamefont {Joshi}, \citenamefont {Lew},\ and\ \citenamefont {Volkas}}]{Foot:1990uf}%
  \BibitemOpen
  \bibfield  {author} {\bibinfo {author} {\bibfnamefont {R.}~\bibnamefont {Foot}}, \bibinfo {author} {\bibfnamefont {G.~C.}\ \bibnamefont {Joshi}}, \bibinfo {author} {\bibfnamefont {H.}~\bibnamefont {Lew}},\ and\ \bibinfo {author} {\bibfnamefont {R.~R.}\ \bibnamefont {Volkas}},\ }\bibfield  {title} {\bibinfo {title} {{Charge quantization in the standard model and some of its extensions}},\ }\href {https://doi.org/10.1142/S0217732390003176} {\bibfield  {journal} {\bibinfo  {journal} {Mod. Phys. Lett. A}\ }\textbf {\bibinfo {volume} {5}},\ \bibinfo {pages} {2721} (\bibinfo {year} {1990})}\BibitemShut {NoStop}%
\bibitem [{\citenamefont {Foot}\ \emph {et~al.}(1994)\citenamefont {Foot}, \citenamefont {He}, \citenamefont {Lew},\ and\ \citenamefont {Volkas}}]{Foot:1994vd}%
  \BibitemOpen
  \bibfield  {author} {\bibinfo {author} {\bibfnamefont {R.}~\bibnamefont {Foot}}, \bibinfo {author} {\bibfnamefont {X.~G.}\ \bibnamefont {He}}, \bibinfo {author} {\bibfnamefont {H.}~\bibnamefont {Lew}},\ and\ \bibinfo {author} {\bibfnamefont {R.~R.}\ \bibnamefont {Volkas}},\ }\bibfield  {title} {\bibinfo {title} {{Model for a light $Z'$ boson}},\ }\href {https://doi.org/10.1103/PhysRevD.50.4571} {\bibfield  {journal} {\bibinfo  {journal} {Phys. Rev. D}\ }\textbf {\bibinfo {volume} {50}},\ \bibinfo {pages} {4571} (\bibinfo {year} {1994})},\ \Eprint {https://arxiv.org/abs/hep-ph/9401250} {arXiv:hep-ph/9401250} \BibitemShut {NoStop}%
\bibitem [{\citenamefont {Langacker}(2009)}]{Langacker:2008yv}%
  \BibitemOpen
  \bibfield  {author} {\bibinfo {author} {\bibfnamefont {P.}~\bibnamefont {Langacker}},\ }\bibfield  {title} {\bibinfo {title} {{The Physics of Heavy $Z^\prime$ Gauge Bosons}},\ }\href {https://doi.org/10.1103/RevModPhys.81.1199} {\bibfield  {journal} {\bibinfo  {journal} {Rev. Mod. Phys.}\ }\textbf {\bibinfo {volume} {81}},\ \bibinfo {pages} {1199} (\bibinfo {year} {2009})},\ \Eprint {https://arxiv.org/abs/0801.1345} {arXiv:0801.1345 [hep-ph]} \BibitemShut {NoStop}%
\bibitem [{\citenamefont {Heeck}\ \emph {et~al.}(2019)\citenamefont {Heeck}, \citenamefont {Lindner}, \citenamefont {Rodejohann},\ and\ \citenamefont {Vogl}}]{Heeck:2018nzc}%
  \BibitemOpen
  \bibfield  {author} {\bibinfo {author} {\bibfnamefont {J.}~\bibnamefont {Heeck}}, \bibinfo {author} {\bibfnamefont {M.}~\bibnamefont {Lindner}}, \bibinfo {author} {\bibfnamefont {W.}~\bibnamefont {Rodejohann}},\ and\ \bibinfo {author} {\bibfnamefont {S.}~\bibnamefont {Vogl}},\ }\bibfield  {title} {\bibinfo {title} {{Non-Standard Neutrino Interactions and Neutral Gauge Bosons}},\ }\href {https://doi.org/10.21468/SciPostPhys.6.3.038} {\bibfield  {journal} {\bibinfo  {journal} {SciPost Phys.}\ }\textbf {\bibinfo {volume} {6}},\ \bibinfo {pages} {038} (\bibinfo {year} {2019})},\ \Eprint {https://arxiv.org/abs/1812.04067} {arXiv:1812.04067 [hep-ph]} \BibitemShut {NoStop}%
\bibitem [{\citenamefont {Garg}\ \emph {et~al.}(2026{\natexlab{a}})\citenamefont {Garg}, \citenamefont {Krishnamoorthi}, \citenamefont {Kumar},\ and\ \citenamefont {Agarwalla}}]{Garg:2026gwx}%
  \BibitemOpen
  \bibfield  {author} {\bibinfo {author} {\bibfnamefont {G.}~\bibnamefont {Garg}}, \bibinfo {author} {\bibfnamefont {J.}~\bibnamefont {Krishnamoorthi}}, \bibinfo {author} {\bibfnamefont {A.}~\bibnamefont {Kumar}},\ and\ \bibinfo {author} {\bibfnamefont {S.~K.}\ \bibnamefont {Agarwalla}},\ }\bibfield  {title} {\bibinfo {title} {{First Constraints on Long-Range Neutrino Interactions using IceCube DeepCore}},\ }\href@noop {} {\  (\bibinfo {year} {2026}{\natexlab{a}})},\ \Eprint {https://arxiv.org/abs/2601.01220} {arXiv:2601.01220 [hep-ph]} \BibitemShut {NoStop}%
\bibitem [{\citenamefont {Wagner}\ \emph {et~al.}(2012)\citenamefont {Wagner}, \citenamefont {Schlamminger}, \citenamefont {Gundlach},\ and\ \citenamefont {Adelberger}}]{Wagner:2012ui}%
  \BibitemOpen
  \bibfield  {author} {\bibinfo {author} {\bibfnamefont {T.~A.}\ \bibnamefont {Wagner}}, \bibinfo {author} {\bibfnamefont {S.}~\bibnamefont {Schlamminger}}, \bibinfo {author} {\bibfnamefont {J.~H.}\ \bibnamefont {Gundlach}},\ and\ \bibinfo {author} {\bibfnamefont {E.~G.}\ \bibnamefont {Adelberger}},\ }\bibfield  {title} {\bibinfo {title} {{Torsion-balance tests of the weak equivalence principle}},\ }\href {https://doi.org/10.1088/0264-9381/29/18/184002} {\bibfield  {journal} {\bibinfo  {journal} {Class. Quant. Grav.}\ }\textbf {\bibinfo {volume} {29}},\ \bibinfo {pages} {184002} (\bibinfo {year} {2012})},\ \Eprint {https://arxiv.org/abs/1207.2442} {arXiv:1207.2442 [gr-qc]} \BibitemShut {NoStop}%
\bibitem [{\citenamefont {Touboul}\ \emph {et~al.}(2022)\citenamefont {Touboul} \emph {et~al.}}]{MICROSCOPE:2022doy}%
  \BibitemOpen
  \bibfield  {author} {\bibinfo {author} {\bibfnamefont {P.}~\bibnamefont {Touboul}} \emph {et~al.} (\bibinfo {collaboration} {MICROSCOPE}),\ }\bibfield  {title} {\bibinfo {title} {{MICROSCOPE Mission: Final Results of the Test of the Equivalence Principle}},\ }\href {https://doi.org/10.1103/PhysRevLett.129.121102} {\bibfield  {journal} {\bibinfo  {journal} {Phys. Rev. Lett.}\ }\textbf {\bibinfo {volume} {129}},\ \bibinfo {pages} {121102} (\bibinfo {year} {2022})},\ \Eprint {https://arxiv.org/abs/2209.15487} {arXiv:2209.15487 [gr-qc]} \BibitemShut {NoStop}%
\bibitem [{\citenamefont {Schlamminger}\ \emph {et~al.}(2008)\citenamefont {Schlamminger}, \citenamefont {Choi}, \citenamefont {Wagner}, \citenamefont {Gundlach},\ and\ \citenamefont {Adelberger}}]{Schlamminger:2007ht}%
  \BibitemOpen
  \bibfield  {author} {\bibinfo {author} {\bibfnamefont {S.}~\bibnamefont {Schlamminger}}, \bibinfo {author} {\bibfnamefont {K.-Y.}\ \bibnamefont {Choi}}, \bibinfo {author} {\bibfnamefont {T.~A.}\ \bibnamefont {Wagner}}, \bibinfo {author} {\bibfnamefont {J.~H.}\ \bibnamefont {Gundlach}},\ and\ \bibinfo {author} {\bibfnamefont {E.~G.}\ \bibnamefont {Adelberger}},\ }\bibfield  {title} {\bibinfo {title} {{Test of the equivalence principle using a rotating torsion balance}},\ }\href {https://doi.org/10.1103/PhysRevLett.100.041101} {\bibfield  {journal} {\bibinfo  {journal} {Phys. Rev. Lett.}\ }\textbf {\bibinfo {volume} {100}},\ \bibinfo {pages} {041101} (\bibinfo {year} {2008})},\ \Eprint {https://arxiv.org/abs/0712.0607} {arXiv:0712.0607 [gr-qc]} \BibitemShut {NoStop}%
\bibitem [{\citenamefont {Adelberger}\ \emph {et~al.}(2009)\citenamefont {Adelberger}, \citenamefont {Gundlach}, \citenamefont {Heckel}, \citenamefont {Hoedl},\ and\ \citenamefont {Schlamminger}}]{Adelberger:2009zz}%
  \BibitemOpen
  \bibfield  {author} {\bibinfo {author} {\bibfnamefont {E.~G.}\ \bibnamefont {Adelberger}}, \bibinfo {author} {\bibfnamefont {J.~H.}\ \bibnamefont {Gundlach}}, \bibinfo {author} {\bibfnamefont {B.~R.}\ \bibnamefont {Heckel}}, \bibinfo {author} {\bibfnamefont {S.}~\bibnamefont {Hoedl}},\ and\ \bibinfo {author} {\bibfnamefont {S.}~\bibnamefont {Schlamminger}},\ }\bibfield  {title} {\bibinfo {title} {{Torsion balance experiments: A low-energy frontier of particle physics}},\ }\href {https://doi.org/10.1016/j.ppnp.2008.08.002} {\bibfield  {journal} {\bibinfo  {journal} {Prog. Part. Nucl. Phys.}\ }\textbf {\bibinfo {volume} {62}},\ \bibinfo {pages} {102} (\bibinfo {year} {2009})}\BibitemShut {NoStop}%
\bibitem [{\citenamefont {Blinnikov}\ \emph {et~al.}(1996)\citenamefont {Blinnikov}, \citenamefont {Dolgov}, \citenamefont {Okun},\ and\ \citenamefont {Voloshin}}]{Blinnikov:1995kp}%
  \BibitemOpen
  \bibfield  {author} {\bibinfo {author} {\bibfnamefont {S.~I.}\ \bibnamefont {Blinnikov}}, \bibinfo {author} {\bibfnamefont {A.~D.}\ \bibnamefont {Dolgov}}, \bibinfo {author} {\bibfnamefont {L.~B.}\ \bibnamefont {Okun}},\ and\ \bibinfo {author} {\bibfnamefont {M.~B.}\ \bibnamefont {Voloshin}},\ }\bibfield  {title} {\bibinfo {title} {{How strong can the coupling of leptonic photons be?}},\ }\href {https://doi.org/10.1016/0550-3213(95)00579-X} {\bibfield  {journal} {\bibinfo  {journal} {Nucl. Phys. B}\ }\textbf {\bibinfo {volume} {458}},\ \bibinfo {pages} {52} (\bibinfo {year} {1996})},\ \Eprint {https://arxiv.org/abs/hep-ph/9505444} {arXiv:hep-ph/9505444} \BibitemShut {NoStop}%
\bibitem [{\citenamefont {Grifols}\ and\ \citenamefont {Mass{\'o}}(2004)}]{Grifols:2003gy}%
  \BibitemOpen
  \bibfield  {author} {\bibinfo {author} {\bibfnamefont {J.~A.}\ \bibnamefont {Grifols}}\ and\ \bibinfo {author} {\bibfnamefont {E.}~\bibnamefont {Mass{\'o}}},\ }\bibfield  {title} {\bibinfo {title} {{Neutrino oscillations in the Sun probe long range leptonic forces}},\ }\href {https://doi.org/10.1016/j.physletb.2003.10.078} {\bibfield  {journal} {\bibinfo  {journal} {Phys. Lett. B}\ }\textbf {\bibinfo {volume} {579}},\ \bibinfo {pages} {123} (\bibinfo {year} {2004})},\ \Eprint {https://arxiv.org/abs/hep-ph/0311141} {arXiv:hep-ph/0311141} \BibitemShut {NoStop}%
\bibitem [{\citenamefont {Bandyopadhyay}\ \emph {et~al.}(2007)\citenamefont {Bandyopadhyay}, \citenamefont {Dighe},\ and\ \citenamefont {Joshipura}}]{Bandyopadhyay:2006uh}%
  \BibitemOpen
  \bibfield  {author} {\bibinfo {author} {\bibfnamefont {A.}~\bibnamefont {Bandyopadhyay}}, \bibinfo {author} {\bibfnamefont {A.}~\bibnamefont {Dighe}},\ and\ \bibinfo {author} {\bibfnamefont {A.~S.}\ \bibnamefont {Joshipura}},\ }\bibfield  {title} {\bibinfo {title} {{Constraints on flavor-dependent long range forces from solar neutrinos and KamLAND}},\ }\href {https://doi.org/10.1103/PhysRevD.75.093005} {\bibfield  {journal} {\bibinfo  {journal} {Phys. Rev. D}\ }\textbf {\bibinfo {volume} {75}},\ \bibinfo {pages} {093005} (\bibinfo {year} {2007})},\ \Eprint {https://arxiv.org/abs/hep-ph/0610263} {arXiv:hep-ph/0610263} \BibitemShut {NoStop}%
\bibitem [{\citenamefont {Gonz{\'a}lez-Garc{\'\i}a}\ \emph {et~al.}(2007)\citenamefont {Gonz{\'a}lez-Garc{\'\i}a}, \citenamefont {de~Holanda}, \citenamefont {Mass{\'o}},\ and\ \citenamefont {Zukanovich~Funchal}}]{Gonzalez-Garcia:2006vic}%
  \BibitemOpen
  \bibfield  {author} {\bibinfo {author} {\bibfnamefont {M.~C.}\ \bibnamefont {Gonz{\'a}lez-Garc{\'\i}a}}, \bibinfo {author} {\bibfnamefont {P.~C.}\ \bibnamefont {de~Holanda}}, \bibinfo {author} {\bibfnamefont {E.}~\bibnamefont {Mass{\'o}}},\ and\ \bibinfo {author} {\bibfnamefont {R.}~\bibnamefont {Zukanovich~Funchal}},\ }\bibfield  {title} {\bibinfo {title} {{Probing long-range leptonic forces with solar and reactor neutrinos}},\ }\href {https://doi.org/10.1088/1475-7516/2007/01/005} {\bibfield  {journal} {\bibinfo  {journal} {JCAP}\ }\textbf {\bibinfo {volume} {01}}\bibfield  {number} {\bibinfo  {number} { (2007)},\ \bibinfo {pages} {005}},\ }\Eprint {https://arxiv.org/abs/hep-ph/0609094} {arXiv:hep-ph/0609094} \BibitemShut {NoStop}%
\bibitem [{\citenamefont {Joshipura}\ and\ \citenamefont {Mohanty}(2004)}]{Joshipura:2003jh}%
  \BibitemOpen
  \bibfield  {author} {\bibinfo {author} {\bibfnamefont {A.~S.}\ \bibnamefont {Joshipura}}\ and\ \bibinfo {author} {\bibfnamefont {S.}~\bibnamefont {Mohanty}},\ }\bibfield  {title} {\bibinfo {title} {{Constraints on flavor dependent long range forces from atmospheric neutrino observations at Super-Kamiokande}},\ }\href {https://doi.org/10.1016/j.physletb.2004.01.057} {\bibfield  {journal} {\bibinfo  {journal} {Phys. Lett. B}\ }\textbf {\bibinfo {volume} {584}},\ \bibinfo {pages} {103} (\bibinfo {year} {2004})},\ \Eprint {https://arxiv.org/abs/hep-ph/0310210} {arXiv:hep-ph/0310210} \BibitemShut {NoStop}%
\bibitem [{\citenamefont {Samanta}(2011)}]{Samanta:2010zh}%
  \BibitemOpen
  \bibfield  {author} {\bibinfo {author} {\bibfnamefont {A.}~\bibnamefont {Samanta}},\ }\bibfield  {title} {\bibinfo {title} {{Long-range Forces: Atmospheric Neutrino Oscillation at a magnetized Detector}},\ }\href {https://doi.org/10.1088/1475-7516/2011/09/010} {\bibfield  {journal} {\bibinfo  {journal} {JCAP}\ }\textbf {\bibinfo {volume} {09}}\bibfield  {number} {\bibinfo  {number} { (2011)},\ \bibinfo {pages} {010}},\ }\Eprint {https://arxiv.org/abs/1001.5344} {arXiv:1001.5344 [hep-ph]} \BibitemShut {NoStop}%
\bibitem [{\citenamefont {Khatun}\ \emph {et~al.}(2018)\citenamefont {Khatun}, \citenamefont {Thakore},\ and\ \citenamefont {Agarwalla}}]{Khatun:2018lzs}%
  \BibitemOpen
  \bibfield  {author} {\bibinfo {author} {\bibfnamefont {A.}~\bibnamefont {Khatun}}, \bibinfo {author} {\bibfnamefont {T.}~\bibnamefont {Thakore}},\ and\ \bibinfo {author} {\bibfnamefont {S.~K.}\ \bibnamefont {Agarwalla}},\ }\bibfield  {title} {\bibinfo {title} {{Can INO be Sensitive to Flavor-Dependent Long-Range Forces?}},\ }\href {https://doi.org/10.1007/JHEP04(2018)023} {\bibfield  {journal} {\bibinfo  {journal} {JHEP}\ }\textbf {\bibinfo {volume} {04}}\bibfield  {number} {\bibinfo  {number} { (2018)},\ \bibinfo {pages} {023}},\ }\Eprint {https://arxiv.org/abs/1801.00949} {arXiv:1801.00949 [hep-ph]} \BibitemShut {NoStop}%
\bibitem [{\citenamefont {Heeck}\ and\ \citenamefont {Rodejohann}(2011)}]{Heeck:2010pg}%
  \BibitemOpen
  \bibfield  {author} {\bibinfo {author} {\bibfnamefont {J.}~\bibnamefont {Heeck}}\ and\ \bibinfo {author} {\bibfnamefont {W.}~\bibnamefont {Rodejohann}},\ }\bibfield  {title} {\bibinfo {title} {{Gauged $L_\mu - L_\tau$ and different Muon Neutrino and Anti-Neutrino Oscillations: MINOS and beyond}},\ }\href {https://doi.org/10.1088/0954-3899/38/8/085005} {\bibfield  {journal} {\bibinfo  {journal} {J. Phys. G}\ }\textbf {\bibinfo {volume} {38}},\ \bibinfo {pages} {085005} (\bibinfo {year} {2011})},\ \Eprint {https://arxiv.org/abs/1007.2655} {arXiv:1007.2655 [hep-ph]} \BibitemShut {NoStop}%
\bibitem [{\citenamefont {Davoudiasl}\ \emph {et~al.}(2011)\citenamefont {Davoudiasl}, \citenamefont {Lee},\ and\ \citenamefont {Marciano}}]{Davoudiasl:2011sz}%
  \BibitemOpen
  \bibfield  {author} {\bibinfo {author} {\bibfnamefont {H.}~\bibnamefont {Davoudiasl}}, \bibinfo {author} {\bibfnamefont {H.-S.}\ \bibnamefont {Lee}},\ and\ \bibinfo {author} {\bibfnamefont {W.~J.}\ \bibnamefont {Marciano}},\ }\bibfield  {title} {\bibinfo {title} {{Long-Range Lepton Flavor Interactions and Neutrino Oscillations}},\ }\href {https://doi.org/10.1103/PhysRevD.84.013009} {\bibfield  {journal} {\bibinfo  {journal} {Phys. Rev. D}\ }\textbf {\bibinfo {volume} {84}},\ \bibinfo {pages} {013009} (\bibinfo {year} {2011})},\ \Eprint {https://arxiv.org/abs/1102.5352} {arXiv:1102.5352 [hep-ph]} \BibitemShut {NoStop}%
\bibitem [{\citenamefont {Chatterjee}\ \emph {et~al.}(2015)\citenamefont {Chatterjee}, \citenamefont {Dasgupta},\ and\ \citenamefont {Agarwalla}}]{Chatterjee:2015gta}%
  \BibitemOpen
  \bibfield  {author} {\bibinfo {author} {\bibfnamefont {S.~S.}\ \bibnamefont {Chatterjee}}, \bibinfo {author} {\bibfnamefont {A.}~\bibnamefont {Dasgupta}},\ and\ \bibinfo {author} {\bibfnamefont {S.~K.}\ \bibnamefont {Agarwalla}},\ }\bibfield  {title} {\bibinfo {title} {{Exploring Flavor-Dependent Long-Range Forces in Long-Baseline Neutrino Oscillation Experiments}},\ }\href {https://doi.org/10.1007/JHEP12(2015)167} {\bibfield  {journal} {\bibinfo  {journal} {JHEP}\ }\textbf {\bibinfo {volume} {12}}\bibfield  {number} {\bibinfo  {number} { (2015)},\ \bibinfo {pages} {167}},\ }\Eprint {https://arxiv.org/abs/1509.03517} {arXiv:1509.03517 [hep-ph]} \BibitemShut {NoStop}%
\bibitem [{\citenamefont {Singh}\ \emph {et~al.}(2023)\citenamefont {Singh}, \citenamefont {Bustamante},\ and\ \citenamefont {Agarwalla}}]{Singh:2023nek}%
  \BibitemOpen
  \bibfield  {author} {\bibinfo {author} {\bibfnamefont {M.}~\bibnamefont {Singh}}, \bibinfo {author} {\bibfnamefont {M.}~\bibnamefont {Bustamante}},\ and\ \bibinfo {author} {\bibfnamefont {S.~K.}\ \bibnamefont {Agarwalla}},\ }\bibfield  {title} {\bibinfo {title} {{Flavor-dependent long-range neutrino interactions in DUNE {\&} T2HK: alone they constrain, together they discover}},\ }\href {https://doi.org/10.1007/JHEP08(2023)101} {\bibfield  {journal} {\bibinfo  {journal} {JHEP}\ }\textbf {\bibinfo {volume} {08}}\bibfield  {number} {\bibinfo  {number} { (2023)},\ \bibinfo {pages} {101}},\ }\Eprint {https://arxiv.org/abs/2305.05184} {arXiv:2305.05184 [hep-ph]} \BibitemShut {NoStop}%
\bibitem [{\citenamefont {Agarwalla}\ \emph {et~al.}(2024)\citenamefont {Agarwalla}, \citenamefont {Bustamante}, \citenamefont {Singh},\ and\ \citenamefont {Swain}}]{Agarwalla:2024ylc}%
  \BibitemOpen
  \bibfield  {author} {\bibinfo {author} {\bibfnamefont {S.~K.}\ \bibnamefont {Agarwalla}}, \bibinfo {author} {\bibfnamefont {M.}~\bibnamefont {Bustamante}}, \bibinfo {author} {\bibfnamefont {M.}~\bibnamefont {Singh}},\ and\ \bibinfo {author} {\bibfnamefont {P.}~\bibnamefont {Swain}},\ }\bibfield  {title} {\bibinfo {title} {{A plethora of long-range neutrino interactions probed by DUNE and T2HK}},\ }\href {https://doi.org/10.1007/JHEP09(2024)055} {\bibfield  {journal} {\bibinfo  {journal} {JHEP}\ }\textbf {\bibinfo {volume} {09}}\bibfield  {number} {\bibinfo  {number} { (2024)},\ \bibinfo {pages} {055}},\ }\Eprint {https://arxiv.org/abs/2404.02775} {arXiv:2404.02775 [hep-ph]} \BibitemShut {NoStop}%
\bibitem [{\citenamefont {Bustamante}\ and\ \citenamefont {Agarwalla}(2019)}]{Bustamante:2018mzu}%
  \BibitemOpen
  \bibfield  {author} {\bibinfo {author} {\bibfnamefont {M.}~\bibnamefont {Bustamante}}\ and\ \bibinfo {author} {\bibfnamefont {S.~K.}\ \bibnamefont {Agarwalla}},\ }\bibfield  {title} {\bibinfo {title} {{Universe's Worth of Electrons to Probe Long-Range Interactions of High-Energy Astrophysical Neutrinos}},\ }\href {https://doi.org/10.1103/PhysRevLett.122.061103} {\bibfield  {journal} {\bibinfo  {journal} {Phys. Rev. Lett.}\ }\textbf {\bibinfo {volume} {122}},\ \bibinfo {pages} {061103} (\bibinfo {year} {2019})},\ \Eprint {https://arxiv.org/abs/1808.02042} {arXiv:1808.02042 [astro-ph.HE]} \BibitemShut {NoStop}%
\bibitem [{\citenamefont {Agarwalla}\ \emph {et~al.}(2023)\citenamefont {Agarwalla}, \citenamefont {Bustamante}, \citenamefont {Das},\ and\ \citenamefont {Narang}}]{Agarwalla:2023sng}%
  \BibitemOpen
  \bibfield  {author} {\bibinfo {author} {\bibfnamefont {S.~K.}\ \bibnamefont {Agarwalla}}, \bibinfo {author} {\bibfnamefont {M.}~\bibnamefont {Bustamante}}, \bibinfo {author} {\bibfnamefont {S.}~\bibnamefont {Das}},\ and\ \bibinfo {author} {\bibfnamefont {A.}~\bibnamefont {Narang}},\ }\bibfield  {title} {\bibinfo {title} {{Present and future constraints on flavor-dependent long-range interactions of high-energy astrophysical neutrinos}},\ }\href {https://doi.org/10.1007/JHEP08(2023)113} {\bibfield  {journal} {\bibinfo  {journal} {JHEP}\ }\textbf {\bibinfo {volume} {08}}\bibfield  {number} {\bibinfo  {number} { (2023)},\ \bibinfo {pages} {113}},\ }\Eprint {https://arxiv.org/abs/2305.03675} {arXiv:2305.03675 [hep-ph]} \BibitemShut {NoStop}%
\bibitem [{\citenamefont {Abbasi}\ \emph {et~al.}(2012)\citenamefont {Abbasi} \emph {et~al.}}]{IceCube:2011ucd}%
  \BibitemOpen
  \bibfield  {author} {\bibinfo {author} {\bibfnamefont {R.}~\bibnamefont {Abbasi}} \emph {et~al.} (\bibinfo {collaboration} {IceCube}),\ }\bibfield  {title} {\bibinfo {title} {{The Design and Performance of IceCube DeepCore}},\ }\href {https://doi.org/10.1016/j.astropartphys.2012.01.004} {\bibfield  {journal} {\bibinfo  {journal} {Astropart. Phys.}\ }\textbf {\bibinfo {volume} {35}},\ \bibinfo {pages} {615} (\bibinfo {year} {2012})},\ \Eprint {https://arxiv.org/abs/1109.6096} {arXiv:1109.6096 [astro-ph.IM]} \BibitemShut {NoStop}%
\bibitem [{\citenamefont {Abbasi}\ \emph {et~al.}(2023)\citenamefont {Abbasi} \emph {et~al.}}]{IceCubeCollaboration:2023wtb}%
  \BibitemOpen
  \bibfield  {author} {\bibinfo {author} {\bibfnamefont {R.}~\bibnamefont {Abbasi}} \emph {et~al.} (\bibinfo {collaboration} {IceCube}),\ }\bibfield  {title} {\bibinfo {title} {{Measurement of atmospheric neutrino mixing with improved IceCube DeepCore calibration and data processing}},\ }\href {https://doi.org/10.1103/PhysRevD.108.012014} {\bibfield  {journal} {\bibinfo  {journal} {Phys. Rev. D}\ }\textbf {\bibinfo {volume} {108}},\ \bibinfo {pages} {012014} (\bibinfo {year} {2023})},\ \Eprint {https://arxiv.org/abs/2304.12236} {arXiv:2304.12236 [hep-ex]} \BibitemShut {NoStop}%
\bibitem [{\citenamefont {Garg}\ \emph {et~al.}(2026{\natexlab{b}})\citenamefont {Garg}, \citenamefont {Krishnamoorthi}, \citenamefont {Kumar},\ and\ \citenamefont {Agarwalla}}]{Garg:2026rmy}%
  \BibitemOpen
  \bibfield  {author} {\bibinfo {author} {\bibfnamefont {G.}~\bibnamefont {Garg}}, \bibinfo {author} {\bibfnamefont {J.}~\bibnamefont {Krishnamoorthi}}, \bibinfo {author} {\bibfnamefont {A.}~\bibnamefont {Kumar}},\ and\ \bibinfo {author} {\bibfnamefont {S.~K.}\ \bibnamefont {Agarwalla}},\ }\bibfield  {title} {\bibinfo {title} {{Constraints on long-range neutrino interactions from a variety of $U(1)^\prime$ symmetries using atmospheric neutrinos at IceCube DeepCore}},\ }\href@noop {} {\  (\bibinfo {year} {2026}{\natexlab{b}})},\ \Eprint {https://arxiv.org/abs/2606.03443} {arXiv:2606.03443 [hep-ph]} \BibitemShut {NoStop}%
\bibitem [{\citenamefont {Coloma}\ \emph {et~al.}(2021)\citenamefont {Coloma}, \citenamefont {Gonz{\'a}lez-Garc{\'\i}a},\ and\ \citenamefont {Maltoni}}]{Coloma:2020gfv}%
  \BibitemOpen
  \bibfield  {author} {\bibinfo {author} {\bibfnamefont {P.}~\bibnamefont {Coloma}}, \bibinfo {author} {\bibfnamefont {M.~C.}\ \bibnamefont {Gonz{\'a}lez-Garc{\'\i}a}},\ and\ \bibinfo {author} {\bibfnamefont {M.}~\bibnamefont {Maltoni}},\ }\bibfield  {title} {\bibinfo {title} {{Neutrino oscillation constraints on $U(1)'$ models: from non-standard interactions to long-range forces}},\ }\href {https://doi.org/10.1007/JHEP01(2021)114} {\bibfield  {journal} {\bibinfo  {journal} {JHEP}\ }\textbf {\bibinfo {volume} {01}}\bibfield  {number} {\bibinfo  {number} { (2021)},\ \bibinfo {pages} {114}},\ }\bibinfo {note} {[Erratum: JHEP 11, 115 (2022)]},\ \Eprint {https://arxiv.org/abs/2009.14220} {arXiv:2009.14220 [hep-ph]} \BibitemShut {NoStop}%
\bibitem [{\citenamefont {Smirnov}\ and\ \citenamefont {Xu}(2019)}]{Smirnov:2019cae}%
  \BibitemOpen
  \bibfield  {author} {\bibinfo {author} {\bibfnamefont {A.~Y.}\ \bibnamefont {Smirnov}}\ and\ \bibinfo {author} {\bibfnamefont {X.-J.}\ \bibnamefont {Xu}},\ }\bibfield  {title} {\bibinfo {title} {{Wolfenstein potentials for neutrinos induced by ultra-light mediators}},\ }\href {https://doi.org/10.1007/JHEP12(2019)046} {\bibfield  {journal} {\bibinfo  {journal} {JHEP}\ }\textbf {\bibinfo {volume} {12}}\bibfield  {number} {\bibinfo  {number} { (2019)},\ \bibinfo {pages} {046}},\ }\Eprint {https://arxiv.org/abs/1909.07505} {arXiv:1909.07505 [hep-ph]} \BibitemShut {NoStop}%
\bibitem [{\citenamefont {Wise}\ and\ \citenamefont {Zhang}(2018)}]{Wise:2018rnb}%
  \BibitemOpen
  \bibfield  {author} {\bibinfo {author} {\bibfnamefont {M.~B.}\ \bibnamefont {Wise}}\ and\ \bibinfo {author} {\bibfnamefont {Y.}~\bibnamefont {Zhang}},\ }\bibfield  {title} {\bibinfo {title} {{Lepton Flavorful Fifth Force and Depth-dependent Neutrino Matter Interactions}},\ }\href {https://doi.org/10.1007/JHEP06(2018)053} {\bibfield  {journal} {\bibinfo  {journal} {JHEP}\ }\textbf {\bibinfo {volume} {06}}\bibfield  {number} {\bibinfo  {number} { (2018)},\ \bibinfo {pages} {053}},\ }\Eprint {https://arxiv.org/abs/1803.00591} {arXiv:1803.00591 [hep-ph]} \BibitemShut {NoStop}%
\bibitem [{\citenamefont {Araki}\ \emph {et~al.}(2012)\citenamefont {Araki}, \citenamefont {Heeck},\ and\ \citenamefont {Kubo}}]{Araki:2012ip}%
  \BibitemOpen
  \bibfield  {author} {\bibinfo {author} {\bibfnamefont {T.}~\bibnamefont {Araki}}, \bibinfo {author} {\bibfnamefont {J.}~\bibnamefont {Heeck}},\ and\ \bibinfo {author} {\bibfnamefont {J.}~\bibnamefont {Kubo}},\ }\bibfield  {title} {\bibinfo {title} {{Vanishing Minors in the Neutrino Mass Matrix from Abelian Gauge Symmetries}},\ }\href {https://doi.org/10.1007/JHEP07(2012)083} {\bibfield  {journal} {\bibinfo  {journal} {JHEP}\ }\textbf {\bibinfo {volume} {07}}\bibfield  {number} {\bibinfo  {number} { (2012)},\ \bibinfo {pages} {083}},\ }\Eprint {https://arxiv.org/abs/1203.4951} {arXiv:1203.4951 [hep-ph]} \BibitemShut {NoStop}%
\bibitem [{\citenamefont {Allanach}\ \emph {et~al.}(2019)\citenamefont {Allanach}, \citenamefont {Davighi},\ and\ \citenamefont {Melville}}]{Allanach:2018vjg}%
  \BibitemOpen
  \bibfield  {author} {\bibinfo {author} {\bibfnamefont {B.~C.}\ \bibnamefont {Allanach}}, \bibinfo {author} {\bibfnamefont {J.}~\bibnamefont {Davighi}},\ and\ \bibinfo {author} {\bibfnamefont {S.}~\bibnamefont {Melville}},\ }\bibfield  {title} {\bibinfo {title} {{An Anomaly-free Atlas: charting the space of flavour-dependent gauged $U(1)$ extensions of the Standard Model}},\ }\href {https://doi.org/10.1007/JHEP02(2019)082} {\bibfield  {journal} {\bibinfo  {journal} {JHEP}\ }\textbf {\bibinfo {volume} {02}}\bibfield  {number} {\bibinfo  {number} { (2019)},\ \bibinfo {pages} {082}},\ }\bibinfo {note} {[Erratum: JHEP 08, 064 (2019)]},\ \Eprint {https://arxiv.org/abs/1812.04602} {arXiv:1812.04602 [hep-ph]} \BibitemShut {NoStop}%
\bibitem [{\citenamefont {de~la Vega}\ \emph {et~al.}(2021)\citenamefont {de~la Vega}, \citenamefont {Flores}, \citenamefont {Nath},\ and\ \citenamefont {Peinado}}]{delaVega:2021wpx}%
  \BibitemOpen
  \bibfield  {author} {\bibinfo {author} {\bibfnamefont {L.~M.~G.}\ \bibnamefont {de~la Vega}}, \bibinfo {author} {\bibfnamefont {L.~J.}\ \bibnamefont {Flores}}, \bibinfo {author} {\bibfnamefont {N.}~\bibnamefont {Nath}},\ and\ \bibinfo {author} {\bibfnamefont {E.}~\bibnamefont {Peinado}},\ }\bibfield  {title} {\bibinfo {title} {{Complementarity between dark matter direct searches and CE$\nu$NS experiments in $U(1)'$ models}},\ }\href {https://doi.org/10.1007/JHEP09(2021)146} {\bibfield  {journal} {\bibinfo  {journal} {JHEP}\ }\textbf {\bibinfo {volume} {09}}\bibfield  {number} {\bibinfo  {number} { (2021)},\ \bibinfo {pages} {146}},\ }\Eprint {https://arxiv.org/abs/2107.04037} {arXiv:2107.04037 [hep-ph]} \BibitemShut {NoStop}%
\bibitem [{\citenamefont {Ma}(1998)}]{Ma:1997nq}%
  \BibitemOpen
  \bibfield  {author} {\bibinfo {author} {\bibfnamefont {E.}~\bibnamefont {Ma}},\ }\bibfield  {title} {\bibinfo {title} {{Gauged $B - 3L_\tau$ and radiative neutrino masses}},\ }\href {https://doi.org/10.1016/S0370-2693(98)00599-1} {\bibfield  {journal} {\bibinfo  {journal} {Phys. Lett. B}\ }\textbf {\bibinfo {volume} {433}},\ \bibinfo {pages} {74} (\bibinfo {year} {1998})},\ \Eprint {https://arxiv.org/abs/hep-ph/9709474} {arXiv:hep-ph/9709474} \BibitemShut {NoStop}%
\bibitem [{\citenamefont {Lee}\ and\ \citenamefont {Ma}(2010)}]{Lee:2010hf}%
  \BibitemOpen
  \bibfield  {author} {\bibinfo {author} {\bibfnamefont {H.-S.}\ \bibnamefont {Lee}}\ and\ \bibinfo {author} {\bibfnamefont {E.}~\bibnamefont {Ma}},\ }\bibfield  {title} {\bibinfo {title} {{Gauged $B-x_i L$ origin of $R$ Parity and its implications}},\ }\href {https://doi.org/10.1016/j.physletb.2010.04.032} {\bibfield  {journal} {\bibinfo  {journal} {Phys. Lett. B}\ }\textbf {\bibinfo {volume} {688}},\ \bibinfo {pages} {319} (\bibinfo {year} {2010})},\ \Eprint {https://arxiv.org/abs/1001.0768} {arXiv:1001.0768 [hep-ph]} \BibitemShut {NoStop}%
\bibitem [{\citenamefont {Kownacki}\ \emph {et~al.}(2017)\citenamefont {Kownacki}, \citenamefont {Ma}, \citenamefont {Pollard},\ and\ \citenamefont {Zakeri}}]{Kownacki:2016pmx}%
  \BibitemOpen
  \bibfield  {author} {\bibinfo {author} {\bibfnamefont {C.}~\bibnamefont {Kownacki}}, \bibinfo {author} {\bibfnamefont {E.}~\bibnamefont {Ma}}, \bibinfo {author} {\bibfnamefont {N.}~\bibnamefont {Pollard}},\ and\ \bibinfo {author} {\bibfnamefont {M.}~\bibnamefont {Zakeri}},\ }\bibfield  {title} {\bibinfo {title} {{Generalized Gauge $U(1)$ Family Symmetry for Quarks and Leptons}},\ }\href {https://doi.org/10.1016/j.physletb.2017.01.013} {\bibfield  {journal} {\bibinfo  {journal} {Phys. Lett. B}\ }\textbf {\bibinfo {volume} {766}},\ \bibinfo {pages} {149} (\bibinfo {year} {2017})},\ \Eprint {https://arxiv.org/abs/1611.05017} {arXiv:1611.05017 [hep-ph]} \BibitemShut {NoStop}%
\bibitem [{\citenamefont {Farzan}(2015)}]{Farzan:2015doa}%
  \BibitemOpen
  \bibfield  {author} {\bibinfo {author} {\bibfnamefont {Y.}~\bibnamefont {Farzan}},\ }\bibfield  {title} {\bibinfo {title} {{A model for large non-standard interactions of neutrinos leading to the LMA-Dark solution}},\ }\href {https://doi.org/10.1016/j.physletb.2015.07.015} {\bibfield  {journal} {\bibinfo  {journal} {Phys. Lett. B}\ }\textbf {\bibinfo {volume} {748}},\ \bibinfo {pages} {311} (\bibinfo {year} {2015})},\ \Eprint {https://arxiv.org/abs/1505.06906} {arXiv:1505.06906 [hep-ph]} \BibitemShut {NoStop}%
\bibitem [{\citenamefont {Holdom}(1986)}]{Holdom:1985ag}%
  \BibitemOpen
  \bibfield  {author} {\bibinfo {author} {\bibfnamefont {B.}~\bibnamefont {Holdom}},\ }\bibfield  {title} {\bibinfo {title} {{Two $U(1)$'s and $\epsilon$ Charge Shifts}},\ }\href {https://doi.org/10.1016/0370-2693(86)91377-8} {\bibfield  {journal} {\bibinfo  {journal} {Phys. Lett. B}\ }\textbf {\bibinfo {volume} {166}},\ \bibinfo {pages} {196} (\bibinfo {year} {1986})}\BibitemShut {NoStop}%
\bibitem [{\citenamefont {Babu}\ \emph {et~al.}(1998)\citenamefont {Babu}, \citenamefont {Kolda},\ and\ \citenamefont {March-Russell}}]{Babu:1997st}%
  \BibitemOpen
  \bibfield  {author} {\bibinfo {author} {\bibfnamefont {K.~S.}\ \bibnamefont {Babu}}, \bibinfo {author} {\bibfnamefont {C.~F.}\ \bibnamefont {Kolda}},\ and\ \bibinfo {author} {\bibfnamefont {J.}~\bibnamefont {March-Russell}},\ }\bibfield  {title} {\bibinfo {title} {{Implications of generalized $Z$-$Z'$ mixing}},\ }\href {https://doi.org/10.1103/PhysRevD.57.6788} {\bibfield  {journal} {\bibinfo  {journal} {Phys. Rev. D}\ }\textbf {\bibinfo {volume} {57}},\ \bibinfo {pages} {6788} (\bibinfo {year} {1998})},\ \Eprint {https://arxiv.org/abs/hep-ph/9710441} {arXiv:hep-ph/9710441} \BibitemShut {NoStop}%
\bibitem [{\citenamefont {Joshipura}\ \emph {et~al.}(2020)\citenamefont {Joshipura}, \citenamefont {Mahajan},\ and\ \citenamefont {Patel}}]{Joshipura:2019qxz}%
  \BibitemOpen
  \bibfield  {author} {\bibinfo {author} {\bibfnamefont {A.~S.}\ \bibnamefont {Joshipura}}, \bibinfo {author} {\bibfnamefont {N.}~\bibnamefont {Mahajan}},\ and\ \bibinfo {author} {\bibfnamefont {K.~M.}\ \bibnamefont {Patel}},\ }\bibfield  {title} {\bibinfo {title} {{Generalised $\mu$-$\tau$ symmetries and calculable gauge kinetic and mass mixing in $\mathrm{U}{(1)}_{L_{\mu }-{L}_{\tau }}$ models}},\ }\href {https://doi.org/10.1007/JHEP03(2020)001} {\bibfield  {journal} {\bibinfo  {journal} {JHEP}\ }\textbf {\bibinfo {volume} {03}}\bibfield  {number} {\bibinfo  {number} { (2020)},\ \bibinfo {pages} {001}},\ }\Eprint {https://arxiv.org/abs/1909.02331} {arXiv:1909.02331 [hep-ph]} \BibitemShut {NoStop}%
\bibitem [{\citenamefont {Aartsen}\ \emph {et~al.}(2020)\citenamefont {Aartsen} \emph {et~al.}}]{IceCube:2018ikn}%
  \BibitemOpen
  \bibfield  {author} {\bibinfo {author} {\bibfnamefont {M.~G.}\ \bibnamefont {Aartsen}} \emph {et~al.} (\bibinfo {collaboration} {IceCube}),\ }\bibfield  {title} {\bibinfo {title} {{Computational techniques for the analysis of small signals in high-statistics neutrino oscillation experiments}},\ }\href {https://doi.org/10.1016/j.nima.2020.164332} {\bibfield  {journal} {\bibinfo  {journal} {Nucl. Instrum. Meth. A}\ }\textbf {\bibinfo {volume} {977}},\ \bibinfo {pages} {164332} (\bibinfo {year} {2020})},\ \Eprint {https://arxiv.org/abs/1803.05390} {arXiv:1803.05390 [physics.data-an]} \BibitemShut {NoStop}%
\bibitem [{\citenamefont {Dziewo{\'n}ski}\ and\ \citenamefont {Anderson}(1981)}]{Dziewonski:1981xy}%
  \BibitemOpen
  \bibfield  {author} {\bibinfo {author} {\bibfnamefont {A.~M.}\ \bibnamefont {Dziewo{\'n}ski}}\ and\ \bibinfo {author} {\bibfnamefont {D.~L.}\ \bibnamefont {Anderson}},\ }\bibfield  {title} {\bibinfo {title} {{Preliminary reference Earth model}},\ }\href {https://doi.org/10.1016/0031-9201(81)90046-7} {\bibfield  {journal} {\bibinfo  {journal} {Phys. Earth Planet. Interiors}\ }\textbf {\bibinfo {volume} {25}},\ \bibinfo {pages} {297} (\bibinfo {year} {1981})}\BibitemShut {NoStop}%
\bibitem [{\citenamefont {Pontecorvo}(1958)}]{Pontecorvo:1957qd}%
  \BibitemOpen
  \bibfield  {author} {\bibinfo {author} {\bibfnamefont {B.}~\bibnamefont {Pontecorvo}},\ }\bibfield  {title} {\bibinfo {title} {{Inverse Beta Processes and Nonconservation of Lepton Charge}},\ }\href@noop {} {\bibfield  {journal} {\bibinfo  {journal} {Sov. Phys. JETP}\ }\textbf {\bibinfo {volume} {7}},\ \bibinfo {pages} {172} (\bibinfo {year} {1958})}\BibitemShut {NoStop}%
\bibitem [{\citenamefont {Pontecorvo}(1968)}]{Pontecorvo:1967fh}%
  \BibitemOpen
  \bibfield  {author} {\bibinfo {author} {\bibfnamefont {B.}~\bibnamefont {Pontecorvo}},\ }\bibfield  {title} {\bibinfo {title} {{Neutrino Experiments and the Problem of Conservation of Leptonic Charge}},\ }\href@noop {} {\bibfield  {journal} {\bibinfo  {journal} {Sov. Phys. JETP}\ }\textbf {\bibinfo {volume} {26}},\ \bibinfo {pages} {984} (\bibinfo {year} {1968})}\BibitemShut {NoStop}%
\bibitem [{\citenamefont {Maki}\ \emph {et~al.}(1962)\citenamefont {Maki}, \citenamefont {Nakagawa},\ and\ \citenamefont {Sakata}}]{Maki:1962mu}%
  \BibitemOpen
  \bibfield  {author} {\bibinfo {author} {\bibfnamefont {Z.}~\bibnamefont {Maki}}, \bibinfo {author} {\bibfnamefont {M.}~\bibnamefont {Nakagawa}},\ and\ \bibinfo {author} {\bibfnamefont {S.}~\bibnamefont {Sakata}},\ }\bibfield  {title} {\bibinfo {title} {{Remarks on the unified model of elementary particles}},\ }\href {https://doi.org/10.1143/PTP.28.870} {\bibfield  {journal} {\bibinfo  {journal} {Prog. Theor. Phys.}\ }\textbf {\bibinfo {volume} {28}},\ \bibinfo {pages} {870} (\bibinfo {year} {1962})}\BibitemShut {NoStop}%
\bibitem [{\citenamefont {Wolfenstein}(1978)}]{Wolfenstein:1977ue}%
  \BibitemOpen
  \bibfield  {author} {\bibinfo {author} {\bibfnamefont {L.}~\bibnamefont {Wolfenstein}},\ }\bibfield  {title} {\bibinfo {title} {{Neutrino Oscillations in Matter}},\ }\href {https://doi.org/10.1103/PhysRevD.17.2369} {\bibfield  {journal} {\bibinfo  {journal} {Phys. Rev. D}\ }\textbf {\bibinfo {volume} {17}},\ \bibinfo {pages} {2369} (\bibinfo {year} {1978})}\BibitemShut {NoStop}%
\bibitem [{\citenamefont {Mikheyev}\ and\ \citenamefont {Smirnov}(1985)}]{Mikheyev:1985zog}%
  \BibitemOpen
  \bibfield  {author} {\bibinfo {author} {\bibfnamefont {S.~P.}\ \bibnamefont {Mikheyev}}\ and\ \bibinfo {author} {\bibfnamefont {A.~Y.}\ \bibnamefont {Smirnov}},\ }\bibfield  {title} {\bibinfo {title} {{Resonance Amplification of Oscillations in Matter and Spectroscopy of Solar Neutrinos}},\ }\href@noop {} {\bibfield  {journal} {\bibinfo  {journal} {Sov. J. Nucl. Phys.}\ }\textbf {\bibinfo {volume} {42}},\ \bibinfo {pages} {913} (\bibinfo {year} {1985})}\BibitemShut {NoStop}%
\bibitem [{\citenamefont {Bustamante}(2026)}]{Bustamante:2026yce}%
  \BibitemOpen
  \bibfield  {author} {\bibinfo {author} {\bibfnamefont {M.}~\bibnamefont {Bustamante}},\ }\bibfield  {title} {\bibinfo {title} {{Magnus: neutrino oscillation probabilities for any Hermitian Hamiltonian, any number of flavors, and any matter profile}},\ }\href@noop {} {\  (\bibinfo {year} {2026})},\ \Eprint {https://arxiv.org/abs/2610.07159} {arXiv:2610.07159 [hep-ph]} \BibitemShut {NoStop}%
\bibitem [{\citenamefont {Gaisser}\ and\ \citenamefont {Honda}(2002)}]{Gaisser:2002jj}%
  \BibitemOpen
  \bibfield  {author} {\bibinfo {author} {\bibfnamefont {T.~K.}\ \bibnamefont {Gaisser}}\ and\ \bibinfo {author} {\bibfnamefont {M.}~\bibnamefont {Honda}},\ }\bibfield  {title} {\bibinfo {title} {{Flux of atmospheric neutrinos}},\ }\href {https://doi.org/10.1146/annurev.nucl.52.050102.090645} {\bibfield  {journal} {\bibinfo  {journal} {Ann. Rev. Nucl. Part. Sci.}\ }\textbf {\bibinfo {volume} {52}},\ \bibinfo {pages} {153} (\bibinfo {year} {2002})},\ \Eprint {https://arxiv.org/abs/hep-ph/0203272} {arXiv:hep-ph/0203272} \BibitemShut {NoStop}%
\bibitem [{\citenamefont {Fedynitch}\ \emph {et~al.}(2015)\citenamefont {Fedynitch}, \citenamefont {Engel}, \citenamefont {Gaisser}, \citenamefont {Riehn},\ and\ \citenamefont {Stanev}}]{Fedynitch:2015zma}%
  \BibitemOpen
  \bibfield  {author} {\bibinfo {author} {\bibfnamefont {A.}~\bibnamefont {Fedynitch}}, \bibinfo {author} {\bibfnamefont {R.}~\bibnamefont {Engel}}, \bibinfo {author} {\bibfnamefont {T.~K.}\ \bibnamefont {Gaisser}}, \bibinfo {author} {\bibfnamefont {F.}~\bibnamefont {Riehn}},\ and\ \bibinfo {author} {\bibfnamefont {T.}~\bibnamefont {Stanev}},\ }\bibfield  {title} {\bibinfo {title} {{Calculation of conventional and prompt lepton fluxes at very high energy}},\ }\href {https://doi.org/10.1051/epjconf/20159908001} {\bibfield  {journal} {\bibinfo  {journal} {EPJ Web Conf.}\ }\textbf {\bibinfo {volume} {99}},\ \bibinfo {pages} {08001} (\bibinfo {year} {2015})},\ \Eprint {https://arxiv.org/abs/1503.00544} {arXiv:1503.00544 [hep-ph]} \BibitemShut {NoStop}%
\bibitem [{\citenamefont {Riehn}\ \emph {et~al.}(2020)\citenamefont {Riehn}, \citenamefont {Engel}, \citenamefont {Fedynitch}, \citenamefont {Gaisser},\ and\ \citenamefont {Stanev}}]{Riehn:2019jet}%
  \BibitemOpen
  \bibfield  {author} {\bibinfo {author} {\bibfnamefont {F.}~\bibnamefont {Riehn}}, \bibinfo {author} {\bibfnamefont {R.}~\bibnamefont {Engel}}, \bibinfo {author} {\bibfnamefont {A.}~\bibnamefont {Fedynitch}}, \bibinfo {author} {\bibfnamefont {T.~K.}\ \bibnamefont {Gaisser}},\ and\ \bibinfo {author} {\bibfnamefont {T.}~\bibnamefont {Stanev}},\ }\bibfield  {title} {\bibinfo {title} {{Hadronic interaction model Sibyll 2.3d and extensive air showers}},\ }\href {https://doi.org/10.1103/PhysRevD.102.063002} {\bibfield  {journal} {\bibinfo  {journal} {Phys. Rev. D}\ }\textbf {\bibinfo {volume} {102}},\ \bibinfo {pages} {063002} (\bibinfo {year} {2020})},\ \Eprint {https://arxiv.org/abs/1912.03300} {arXiv:1912.03300 [hep-ph]} \BibitemShut {NoStop}%
\bibitem [{\citenamefont {Dembinski}\ \emph {et~al.}(2018)\citenamefont {Dembinski}, \citenamefont {Engel}, \citenamefont {Fedynitch}, \citenamefont {Gaisser}, \citenamefont {Riehn},\ and\ \citenamefont {Stanev}}]{Dembinski:2017zsh}%
  \BibitemOpen
  \bibfield  {author} {\bibinfo {author} {\bibfnamefont {H.~P.}\ \bibnamefont {Dembinski}}, \bibinfo {author} {\bibfnamefont {R.}~\bibnamefont {Engel}}, \bibinfo {author} {\bibfnamefont {A.}~\bibnamefont {Fedynitch}}, \bibinfo {author} {\bibfnamefont {T.}~\bibnamefont {Gaisser}}, \bibinfo {author} {\bibfnamefont {F.}~\bibnamefont {Riehn}},\ and\ \bibinfo {author} {\bibfnamefont {T.}~\bibnamefont {Stanev}},\ }\bibfield  {title} {\bibinfo {title} {{Data-driven model of the cosmic-ray flux and mass composition from 10 GeV to $10^{11}$ GeV}},\ }\href {https://doi.org/10.22323/1.301.0533} {\bibfield  {journal} {\bibinfo  {journal} {PoS}\ }\textbf {\bibinfo {volume} {ICRC2017}},\ \bibinfo {pages} {533} (\bibinfo {year} {2018})},\ \Eprint {https://arxiv.org/abs/1711.11432} {arXiv:1711.11432 [astro-ph.HE]} \BibitemShut {NoStop}%
\bibitem [{\citenamefont {Fedynitch}\ \emph {et~al.}(2026)\citenamefont {Fedynitch}, \citenamefont {Fujisue}, \citenamefont {Dembinski},\ and\ \citenamefont {Engel}}]{Fedynitch:2026ugq}%
  \BibitemOpen
  \bibfield  {author} {\bibinfo {author} {\bibfnamefont {A.}~\bibnamefont {Fedynitch}}, \bibinfo {author} {\bibfnamefont {K.}~\bibnamefont {Fujisue}}, \bibinfo {author} {\bibfnamefont {H.}~\bibnamefont {Dembinski}},\ and\ \bibinfo {author} {\bibfnamefont {R.}~\bibnamefont {Engel}},\ }\bibfield  {title} {\bibinfo {title} {{Global Spline Fit: A unified data-driven view of the cosmic-ray spectrum and mass composition from GeV to the highest energies}},\ }\href@noop {} {\  (\bibinfo {year} {2026})},\ \Eprint {https://arxiv.org/abs/2609.32649} {arXiv:2609.32649 [astro-ph.HE]} \BibitemShut {NoStop}%
\bibitem [{\citenamefont {Honda}\ \emph {et~al.}(2015)\citenamefont {Honda}, \citenamefont {Sajjad~Athar}, \citenamefont {Kajita}, \citenamefont {Kasahara},\ and\ \citenamefont {Midorikawa}}]{Honda:2015fha}%
  \BibitemOpen
  \bibfield  {author} {\bibinfo {author} {\bibfnamefont {M.}~\bibnamefont {Honda}}, \bibinfo {author} {\bibfnamefont {M.}~\bibnamefont {Sajjad~Athar}}, \bibinfo {author} {\bibfnamefont {T.}~\bibnamefont {Kajita}}, \bibinfo {author} {\bibfnamefont {K.}~\bibnamefont {Kasahara}},\ and\ \bibinfo {author} {\bibfnamefont {S.}~\bibnamefont {Midorikawa}},\ }\bibfield  {title} {\bibinfo {title} {{Atmospheric neutrino flux calculation using the NRLMSISE-00 atmospheric model}},\ }\href {https://doi.org/10.1103/PhysRevD.92.023004} {\bibfield  {journal} {\bibinfo  {journal} {Phys. Rev. D}\ }\textbf {\bibinfo {volume} {92}},\ \bibinfo {pages} {023004} (\bibinfo {year} {2015})},\ \Eprint {https://arxiv.org/abs/1502.03916} {arXiv:1502.03916 [astro-ph.HE]} \BibitemShut {NoStop}%
\bibitem [{\citenamefont {Cooper-Sarkar}\ \emph {et~al.}(2011)\citenamefont {Cooper-Sarkar}, \citenamefont {Mertsch},\ and\ \citenamefont {Sarkar}}]{Cooper-Sarkar:2011jtt}%
  \BibitemOpen
  \bibfield  {author} {\bibinfo {author} {\bibfnamefont {A.}~\bibnamefont {Cooper-Sarkar}}, \bibinfo {author} {\bibfnamefont {P.}~\bibnamefont {Mertsch}},\ and\ \bibinfo {author} {\bibfnamefont {S.}~\bibnamefont {Sarkar}},\ }\bibfield  {title} {\bibinfo {title} {{The high energy neutrino cross-section in the Standard Model and its uncertainty}},\ }\href {https://doi.org/10.1007/JHEP08(2011)042} {\bibfield  {journal} {\bibinfo  {journal} {JHEP}\ }\textbf {\bibinfo {volume} {08}}\bibfield  {number} {\bibinfo  {number} { (2011)},\ \bibinfo {pages} {042}},\ }\Eprint {https://arxiv.org/abs/1106.3723} {arXiv:1106.3723 [hep-ph]} \BibitemShut {NoStop}%
\bibitem [{\citenamefont {{IceCube Collaboration}}(2025)}]{IceCubeDeepCoreRelease:2025}%
  \BibitemOpen
  \bibfield  {author} {\bibinfo {author} {\bibnamefont {{IceCube Collaboration}}},\ }\href {https://doi.org/10.7910/DVN/B4RITM} {\bibinfo {title} {{Replication Data for: Measurement of atmospheric neutrino mixing with improved IceCube DeepCore calibration and data processing}}},\ \bibinfo {howpublished} {Harvard Dataverse, version 1.0} (\bibinfo {year} {2025})\BibitemShut {NoStop}%
\bibitem [{\citenamefont {Barr}\ \emph {et~al.}(2006)\citenamefont {Barr}, \citenamefont {Gaisser}, \citenamefont {Robbins},\ and\ \citenamefont {Stanev}}]{Barr:2006it}%
  \BibitemOpen
  \bibfield  {author} {\bibinfo {author} {\bibfnamefont {G.~D.}\ \bibnamefont {Barr}}, \bibinfo {author} {\bibfnamefont {T.~K.}\ \bibnamefont {Gaisser}}, \bibinfo {author} {\bibfnamefont {S.}~\bibnamefont {Robbins}},\ and\ \bibinfo {author} {\bibfnamefont {T.}~\bibnamefont {Stanev}},\ }\bibfield  {title} {\bibinfo {title} {{Uncertainties in Atmospheric Neutrino Fluxes}},\ }\href {https://doi.org/10.1103/PhysRevD.74.094009} {\bibfield  {journal} {\bibinfo  {journal} {Phys. Rev. D}\ }\textbf {\bibinfo {volume} {74}},\ \bibinfo {pages} {094009} (\bibinfo {year} {2006})},\ \Eprint {https://arxiv.org/abs/astro-ph/0611266} {arXiv:astro-ph/0611266} \BibitemShut {NoStop}%
\bibitem [{\citenamefont {Andreopoulos}\ \emph {et~al.}(2010)\citenamefont {Andreopoulos} \emph {et~al.}}]{Andreopoulos:2009rq}%
  \BibitemOpen
  \bibfield  {author} {\bibinfo {author} {\bibfnamefont {C.}~\bibnamefont {Andreopoulos}} \emph {et~al.},\ }\bibfield  {title} {\bibinfo {title} {{The GENIE Neutrino Monte Carlo Generator}},\ }\href {https://doi.org/10.1016/j.nima.2009.12.009} {\bibfield  {journal} {\bibinfo  {journal} {Nucl. Instrum. Meth. A}\ }\textbf {\bibinfo {volume} {614}},\ \bibinfo {pages} {87} (\bibinfo {year} {2010})},\ \Eprint {https://arxiv.org/abs/0905.2517} {arXiv:0905.2517 [hep-ph]} \BibitemShut {NoStop}%
\bibitem [{\citenamefont {Esteban}\ \emph {et~al.}(2024)\citenamefont {Esteban}, \citenamefont {Gonz{\'a}lez-Garc{\'\i}a}, \citenamefont {Maltoni}, \citenamefont {Mart{\'\i}nez-Soler}, \citenamefont {Pinheiro},\ and\ \citenamefont {Schwetz}}]{Esteban:2024eli}%
  \BibitemOpen
  \bibfield  {author} {\bibinfo {author} {\bibfnamefont {I.}~\bibnamefont {Esteban}}, \bibinfo {author} {\bibfnamefont {M.~C.}\ \bibnamefont {Gonz{\'a}lez-Garc{\'\i}a}}, \bibinfo {author} {\bibfnamefont {M.}~\bibnamefont {Maltoni}}, \bibinfo {author} {\bibfnamefont {I.}~\bibnamefont {Mart{\'\i}nez-Soler}}, \bibinfo {author} {\bibfnamefont {J.~P.}\ \bibnamefont {Pinheiro}},\ and\ \bibinfo {author} {\bibfnamefont {T.}~\bibnamefont {Schwetz}},\ }\bibfield  {title} {\bibinfo {title} {{NuFit-6.0: updated global analysis of three-flavor neutrino oscillations}},\ }\href {https://doi.org/10.1007/JHEP12(2024)216} {\bibfield  {journal} {\bibinfo  {journal} {JHEP}\ }\textbf {\bibinfo {volume} {12}}\bibfield  {number} {\bibinfo  {number} { (2024)},\ \bibinfo {pages} {216}},\ }\Eprint {https://arxiv.org/abs/2410.05380} {arXiv:2410.05380 [hep-ph]} \BibitemShut {NoStop}%
\bibitem [{NuF()}]{NuFIT61}%
  \BibitemOpen
  \href@noop {} {\bibinfo {title} {{NuFIT 6.1}}},\ \bibinfo {howpublished} {\url{http://www.nu-fit.org}}\BibitemShut {NoStop}%
\bibitem [{\citenamefont {Cowan}\ \emph {et~al.}(2011)\citenamefont {Cowan}, \citenamefont {Cranmer}, \citenamefont {Gross},\ and\ \citenamefont {Vitells}}]{Cowan:2010js}%
  \BibitemOpen
  \bibfield  {author} {\bibinfo {author} {\bibfnamefont {G.}~\bibnamefont {Cowan}}, \bibinfo {author} {\bibfnamefont {K.}~\bibnamefont {Cranmer}}, \bibinfo {author} {\bibfnamefont {E.}~\bibnamefont {Gross}},\ and\ \bibinfo {author} {\bibfnamefont {O.}~\bibnamefont {Vitells}},\ }\bibfield  {title} {\bibinfo {title} {{Asymptotic formulae for likelihood-based tests of new physics}},\ }\href {https://doi.org/10.1140/epjc/s10052-011-1554-0} {\bibfield  {journal} {\bibinfo  {journal} {Eur. Phys. J. C}\ }\textbf {\bibinfo {volume} {71}},\ \bibinfo {pages} {1554} (\bibinfo {year} {2011})},\ \bibinfo {note} {[Erratum: Eur.Phys.J.C 73, 2501 (2013)]},\ \Eprint {https://arxiv.org/abs/1007.1727} {arXiv:1007.1727 [physics.data-an]} \BibitemShut {NoStop}%
\bibitem [{\citenamefont {Wilks}(1938)}]{Wilks:1938dza}%
  \BibitemOpen
  \bibfield  {author} {\bibinfo {author} {\bibfnamefont {S.~S.}\ \bibnamefont {Wilks}},\ }\bibfield  {title} {\bibinfo {title} {{The Large-Sample Distribution of the Likelihood Ratio for Testing Composite Hypotheses}},\ }\href {https://doi.org/10.1214/aoms/1177732360} {\bibfield  {journal} {\bibinfo  {journal} {Annals Math. Statist.}\ }\textbf {\bibinfo {volume} {9}},\ \bibinfo {pages} {60} (\bibinfo {year} {1938})}\BibitemShut {NoStop}%
\bibitem [{\citenamefont {Feldman}\ and\ \citenamefont {Cousins}(1998)}]{Feldman:1997qc}%
  \BibitemOpen
  \bibfield  {author} {\bibinfo {author} {\bibfnamefont {G.~J.}\ \bibnamefont {Feldman}}\ and\ \bibinfo {author} {\bibfnamefont {R.~D.}\ \bibnamefont {Cousins}},\ }\bibfield  {title} {\bibinfo {title} {{A Unified approach to the classical statistical analysis of small signals}},\ }\href {https://doi.org/10.1103/PhysRevD.57.3873} {\bibfield  {journal} {\bibinfo  {journal} {Phys. Rev. D}\ }\textbf {\bibinfo {volume} {57}},\ \bibinfo {pages} {3873} (\bibinfo {year} {1998})},\ \Eprint {https://arxiv.org/abs/physics/9711021} {arXiv:physics/9711021} \BibitemShut {NoStop}%
\bibitem [{\citenamefont {Cardenas}\ and\ \citenamefont {Bustamante}(2026)}]{NuLRIEarthRepo}%
  \BibitemOpen
  \bibfield  {author} {\bibinfo {author} {\bibfnamefont {D.~F.}\ \bibnamefont {Cardenas}}\ and\ \bibinfo {author} {\bibfnamefont {M.}~\bibnamefont {Bustamante}},\ }\href@noop {} {\bibinfo {title} {{nu-lri-earth}}},\ \bibinfo {howpublished} {\url{https://github.com/mbustama/nu-lri-earth}} (\bibinfo {year} {2026})\BibitemShut {NoStop}%
\bibitem [{\citenamefont {Fayet}(2019)}]{Fayet:2018cjy}%
  \BibitemOpen
  \bibfield  {author} {\bibinfo {author} {\bibfnamefont {P.}~\bibnamefont {Fayet}},\ }\bibfield  {title} {\bibinfo {title} {{MICROSCOPE limits on the strength of a new force, with comparisons to gravity and electromagnetism}},\ }\href {https://doi.org/10.1103/PhysRevD.99.055043} {\bibfield  {journal} {\bibinfo  {journal} {Phys. Rev. D}\ }\textbf {\bibinfo {volume} {99}},\ \bibinfo {pages} {055043} (\bibinfo {year} {2019})},\ \Eprint {https://arxiv.org/abs/1809.04991} {arXiv:1809.04991 [hep-ph]} \BibitemShut {NoStop}%
\bibitem [{\citenamefont {Adelberger}\ \emph {et~al.}(2003)\citenamefont {Adelberger}, \citenamefont {Heckel},\ and\ \citenamefont {Nelson}}]{Adelberger:2003zx}%
  \BibitemOpen
  \bibfield  {author} {\bibinfo {author} {\bibfnamefont {E.~G.}\ \bibnamefont {Adelberger}}, \bibinfo {author} {\bibfnamefont {B.~R.}\ \bibnamefont {Heckel}},\ and\ \bibinfo {author} {\bibfnamefont {A.~E.}\ \bibnamefont {Nelson}},\ }\bibfield  {title} {\bibinfo {title} {{Tests of the gravitational inverse square law}},\ }\href {https://doi.org/10.1146/annurev.nucl.53.041002.110503} {\bibfield  {journal} {\bibinfo  {journal} {Ann. Rev. Nucl. Part. Sci.}\ }\textbf {\bibinfo {volume} {53}},\ \bibinfo {pages} {77} (\bibinfo {year} {2003})},\ \Eprint {https://arxiv.org/abs/hep-ph/0307284} {arXiv:hep-ph/0307284} \BibitemShut {NoStop}%
\bibitem [{\citenamefont {Ishihara}(2021)}]{Ishihara:2019aao}%
  \BibitemOpen
  \bibfield  {author} {\bibinfo {author} {\bibfnamefont {A.}~\bibnamefont {Ishihara}} (\bibinfo {collaboration} {IceCube}),\ }\bibfield  {title} {\bibinfo {title} {{The IceCube Upgrade - Design and Science Goals}},\ }\href {https://doi.org/10.22323/1.358.1031} {\bibfield  {journal} {\bibinfo  {journal} {PoS}\ }\textbf {\bibinfo {volume} {ICRC2019}},\ \bibinfo {pages} {1031} (\bibinfo {year} {2021})},\ \Eprint {https://arxiv.org/abs/1908.09441} {arXiv:1908.09441 [astro-ph.HE]} \BibitemShut {NoStop}%
\bibitem [{\citenamefont {Adri{\'a}n-Mart{\'\i}nez}\ \emph {et~al.}(2016)\citenamefont {Adri{\'a}n-Mart{\'\i}nez} \emph {et~al.}}]{KM3Net:2016zxf}%
  \BibitemOpen
  \bibfield  {author} {\bibinfo {author} {\bibfnamefont {S.}~\bibnamefont {Adri{\'a}n-Mart{\'\i}nez}} \emph {et~al.} (\bibinfo {collaboration} {KM3Net}),\ }\bibfield  {title} {\bibinfo {title} {{Letter of intent for KM3NeT 2.0}},\ }\href {https://doi.org/10.1088/0954-3899/43/8/084001} {\bibfield  {journal} {\bibinfo  {journal} {J. Phys. G}\ }\textbf {\bibinfo {volume} {43}},\ \bibinfo {pages} {084001} (\bibinfo {year} {2016})},\ \Eprint {https://arxiv.org/abs/1601.07459} {arXiv:1601.07459 [astro-ph.IM]} \BibitemShut {NoStop}%
\bibitem [{\citenamefont {Abe}\ \emph {et~al.}(2018)\citenamefont {Abe} \emph {et~al.}}]{Hyper-Kamiokande:2018ofw}%
  \BibitemOpen
  \bibfield  {author} {\bibinfo {author} {\bibfnamefont {K.}~\bibnamefont {Abe}} \emph {et~al.} (\bibinfo {collaboration} {Hyper-Kamiokande}),\ }\bibfield  {title} {\bibinfo {title} {{Hyper-Kamiokande Design Report}},\ }\href@noop {} {\  (\bibinfo {year} {2018})},\ \Eprint {https://arxiv.org/abs/1805.04163} {arXiv:1805.04163 [physics.ins-det]} \BibitemShut {NoStop}%
\bibitem [{\citenamefont {Abi}\ \emph {et~al.}(2020)\citenamefont {Abi} \emph {et~al.}}]{DUNE:2020ypp}%
  \BibitemOpen
  \bibfield  {author} {\bibinfo {author} {\bibfnamefont {B.}~\bibnamefont {Abi}} \emph {et~al.} (\bibinfo {collaboration} {DUNE}),\ }\bibfield  {title} {\bibinfo {title} {{Deep Underground Neutrino Experiment (DUNE), Far Detector Technical Design Report, Volume II: DUNE Physics}}\ }\href {https://doi.org/10.2172/1599307} {10.2172/1599307} (\bibinfo {year} {2020}),\ \Eprint {https://arxiv.org/abs/2002.03005} {arXiv:2002.03005 [hep-ex]} \BibitemShut {NoStop}%
\bibitem [{\citenamefont {Picone}\ \emph {et~al.}(2002)\citenamefont {Picone}, \citenamefont {Hedin}, \citenamefont {Drob},\ and\ \citenamefont {Aikin}}]{Picone:2002}%
  \BibitemOpen
  \bibfield  {author} {\bibinfo {author} {\bibfnamefont {J.~M.}\ \bibnamefont {Picone}}, \bibinfo {author} {\bibfnamefont {A.~E.}\ \bibnamefont {Hedin}}, \bibinfo {author} {\bibfnamefont {D.~P.}\ \bibnamefont {Drob}},\ and\ \bibinfo {author} {\bibfnamefont {A.~C.}\ \bibnamefont {Aikin}},\ }\bibfield  {title} {\bibinfo {title} {{NRLMSISE-00 empirical model of the atmosphere: Statistical comparisons and scientific issues}},\ }\href {https://doi.org/10.1029/2002JA009430} {\bibfield  {journal} {\bibinfo  {journal} {J. Geophys. Res. Space Phys.}\ }\textbf {\bibinfo {volume} {107}},\ \bibinfo {pages} {1468} (\bibinfo {year} {2002})}\BibitemShut {NoStop}%
\end{thebibliography}
%

\end{document}